\documentclass[sigconf,screen]{acmart}

\AtBeginDocument{
  \providecommand\BibTeX{{
    \normalfont B\kern-0.5em{\scshape i\kern-0.25em b}\kern-0.8em\TeX}}}

\setcopyright{acmlicensed}
\acmDOI{10.1145/3650212.3652131}
\acmYear{2024}
\copyrightyear{2024}
\acmSubmissionID{issta24main-p712-p}
\acmISBN{979-8-4007-0612-7/24/09}
\acmConference[ISSTA '24]{Proceedings of the 33rd ACM SIGSOFT International Symposium on Software Testing and Analysis}{September 16--20, 2024}{Vienna, Austria}
\acmBooktitle{Proceedings of the 33rd ACM SIGSOFT International Symposium on Software Testing and Analysis (ISSTA '24), September 16--20, 2024, Vienna, Austria}
\received{16-DEC-2023}
\received[accepted]{2024-03-02}

\usepackage{microtype}
\usepackage{multirow}
\usepackage{adjustbox}
\usepackage[linesnumbered,ruled]{algorithm2e}
\usepackage{algpseudocode}
\usepackage{bm}
\usepackage{tabu}
\usepackage{natbib}
\usepackage[utf8]{inputenc}
\usepackage{graphicx}
\usepackage{float}
\usepackage{booktabs}
\usepackage{arydshln}
\usepackage{xparse}
\usepackage{longtable}
\usepackage[para,online,flushleft]{threeparttable}
\usepackage{dashrule}
\usepackage{array}
\usepackage{makecell}
\usepackage{subfigure}
\usepackage{adjustbox}
\usepackage{amsmath,amsfonts}
\usepackage{amsthm}
\usepackage{array}
\usepackage{paralist}
\usepackage{setspace}
\usepackage[normalem]{ulem}
\usepackage{url}
\usepackage[most]{tcolorbox}
\usepackage{soul}
\usepackage{hyperref}
\usepackage{romannum}
\usepackage{tabu}
\usepackage{tablefootnote}
\usepackage{tabularx}
\usepackage{textcomp}
\usepackage{threeparttable}
\usepackage{xcolor}
\usepackage{xspace}
\usepackage{enumitem}
\setitemize{leftmargin=*}

\usepackage[]{collab}

\usepackage{cleveref}
\crefname{section}{§}{§§}
\Crefname{section}{§}{§§}

\definecolor{mygreen}{HTML}{719792}
\definecolor{myred}{HTML}{F0616E}
\definecolor{ganyu}{HTML}{2A4597}

\definecolor{background}{HTML}{eaf3ff}
\definecolor{edge}{HTML}{89a7c4}
\newtcolorbox{mybox}{colback=background!35,
colframe=edge,
width=\columnwidth,
arc=2mm,
auto
outer
arc
}

\setlist{leftmargin=18pt,topsep=2pt}
\crefname{section}{§}{§§}
\Crefname{section}{§}{§§}

\newcommand{\frmwk}{{MicroRes}\xspace}
\newcommand{\company}{Huawei Cloud\xspace}

\def\TT{Train-Ticket\xspace}
\def\SN{Social-Network\xspace}
\def\IND{Industry\xspace}
\def\passfail{{\small \texttt{PASS/FAIL}}\xspace}

\makeatletter
\def\@copyrightspace{\relax}
\makeatother

\begin{document}
\title{\frmwk: Versatile Resilience Profiling in Microservices via Degradation Dissemination Indexing}

\author{Tianyi Yang}
\orcid{0000-0003-1492-5197}
\affiliation{
  \institution{The Chinese University of Hong Kong}
  \country{Hong Kong SAR}}
\email{tyyang@cse.cuhk.edu.hk}

\author{Cheryl Lee}
\orcid{0000-0002-0301-3662}
\affiliation{
  \institution{The Chinese University of Hong Kong}
  \country{Hong Kong SAR}}
\email{cheryllee@link.cuhk.edu.hk}

\author{Jiacheng Shen}
\orcid{0009-0000-2855-2403}
\affiliation{
  \institution{The Chinese University of Hong Kong}
  \country{Hong Kong SAR}}
\email{jcshen@cse.cuhk.edu.hk}

\author{Yuxin Su}
\orcid{0000-0002-3338-8561}
\affiliation{
  \institution{Sun Yat-Sen University}
  \city{Zhuhai}
  \country{China}}
\authornote{Yuxin Su is the corresponding author.}
\email{suyx35@mail.sysu.edu.cn}

\author{Cong Feng}
\orcid{0009-0000-5556-4004}
\affiliation{
  \institution{Computing and Networking Innovation Lab, Huawei Cloud Computing Technology Co., Ltd}
  \city{Shenzhen}
  \country{China}}
\email{fengcong5@huawei.com}

\author{Yongqiang Yang}
\orcid{0000-0001-9733-4346}
\affiliation{
  \institution{Computing and Networking Innovation Lab, Huawei Cloud Computing Technology Co., Ltd}
  \city{Shenzhen}
  \country{China}}
\email{yangyongqiang@huawei.com}

\author{Michael R. Lyu}
\orcid{0000-0002-3666-5798}
\affiliation{
  \institution{The Chinese University of Hong Kong}
  \country{Hong Kong SAR}}
\email{lyu@cse.cuhk.edu.hk}

\begin{abstract}
    Microservice resilience, the ability of microservices to recover from failures and continue providing reliable and responsive services, is crucial for cloud vendors.
    However, the current practice relies on manually configured rules specific to a certain microservice system, resulting in labor-intensity and flexibility issues, given the large scale and high dynamics of microservices.
    A more labor-efficient and versatile solution is desired.
    Our insight is that resilient deployment can effectively prevent the dissemination of degradation from system performance metrics to user-aware metrics, and the latter directly affects service quality.
    In other words, failures in a non-resilient deployment can impact both types of metrics, leading to user dissatisfaction.
    With this in mind, we propose \textbf{\frmwk}, the first versatile resilience profiling framework for microservices via degradation dissemination indexing.
    \frmwk first injects failures into microservices and collects available monitoring metrics.
    Then, it ranks the metrics according to their contributions to the overall service degradation. It produces a resilience index by how much the degradation is disseminated from system performance metrics to user-aware metrics.
    Higher degradation dissemination indicates lower resilience.
    We evaluate \frmwk on two open-source and one industrial microservice system.
    The experiments show \frmwk' efficient and effective resilience profiling of microservices.
    We also showcase \frmwk' practical usage in production.
\end{abstract}

\ccsdesc[500]{Computer systems organization~Reliability}
\ccsdesc[500]{Computer systems organization~Maintainability and maintenance}
\ccsdesc[500]{Computer systems organization~Cloud computing}
\ccsdesc[500]{Software and its engineering~Software testing and debugging}

\keywords{Microservices, resilience profiling, fault injection}

\maketitle
\section{Introduction}
\label{sec:intro}

Nowadays, an online service is usually developed as a bunch of fine-grained and independently-managed microservices and then deployed as a microservice system~\cite{microservice-architecture}.
Microservice systems exhibit three prominent attributes~\cite{berkeley-view-cloud}.
First, they are highly decoupled and usually contain many microservices, e.g., Netflix's system has hundreds to thousands of microservices~\cite{Gremlin}.
Second, microservices are dynamic. New features and updates are delivered continuously and frequently.
Last, microservices are specialized. Each microservice only processes a single type of request. Microservices interact with each other and serve users' requests together.

\begin{figure}[ht]
    \centering
    \includegraphics[width=0.8\columnwidth]{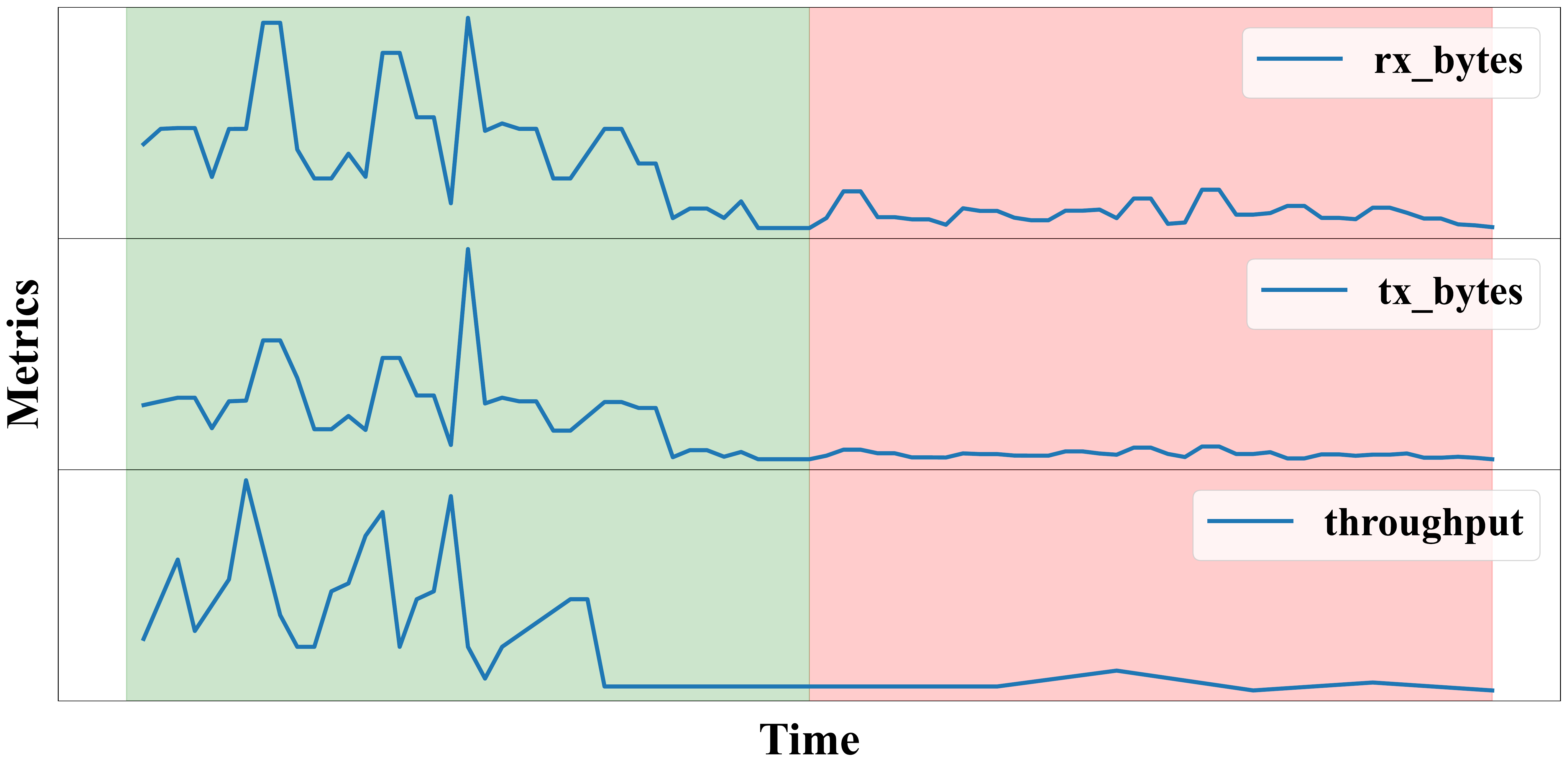}
    \caption{The monitoring metrics during the normal period \textcolor{mygreen}{(the green area)} and the failure injection period \textcolor{myred}{(the red area)}. ``rx\_bytes'' and ``tx\_bytes'' indicate network receive and transmit rate.}
    \label{fig:faulty_metrics}
\end{figure}

Resilience, i.e., the ability to maintain performance at an acceptable level and recover the service back to normal under service failures~\cite{resiliency-definition}, is essentially one of the desired abilities of online services.
Figure~\ref{fig:faulty_metrics} illustrates a non-resilient example by plotting the request throughput of an online service during the \textcolor{mygreen}{normal} and the \textcolor{myred}{faulty} period.
Intuitively, the resilience of the service is low because the failure causes service degradation, reflected by the throughput decrement.
Resilience profiling is thereby crucial as faults and failures are unavoidable~\cite{gray-failure, DBLP:conf/hotos/LiuLMN19} and a resilient system can be commissioned to users by ensuring service reliability.
Without sufficiently high resilience, a new or updated microservice system should not be directly deployed in the production environment.

The current practice~\cite{techtarget-blog} for resilience profiling is to set resilience rules manually, including the concerned failure types, the metrics to monitor, the measure of degradation, and the criteria for passing or failing the tests.
However, such a method is highly time-consuming and labor-intensive, all the while lacking flexibility to adapt to different microservice systems.

First, manual rule identification relies heavily on domain expertise to define the rules that can represent the degradation caused by failures.
Defining proper rules is very burdensome because
1) the number of microservices is usually huge (up to tens of thousands), and so are their failures since microservices are highly decoupled~\cite{AID};
2) the dynamism of microservices requires frequent updates of the rules~\cite{berkeley-view-cloud}.
In \company, it usually takes two man-months of discussion before the test engineers reach a consensus on the rules according to our survey, and the update requirement even worsens this situation.

Second, rule-based resilience profiling can not fit in different microservice systems.
The reason is two-fold:
1) Microservices are specialized for different business applications, making the failures and their resulting manifestations manifold~\cite{DBLP:conf/hotos/LiuLMN19}.
2) Fixed \passfail results obtained from resilience test rules fail to discriminate the subtle difference in an online service's resilience when the boundary between ``resilient'' and ``non-resilient'' becomes less absolute.
This is because various refined resilience mechanisms (e.g., circuit breakers, replications, and node auto-scaling) are applied in existing platforms, such as Kubernetes, so the system can be in a ``gray-failure'' status that can not be fully depicted by a few common metrics like mean time to recover (MTTR) used in the test rules.

An intuitive idea to mitigate the two issues is to propose a versatile resilience profiling technique with smooth criteria.
However, designing such an approach is non-trivial.
The critical challenge is how to determine to what extent a microservice system is resilient.
To address the challenge, we investigate the failure impact on two deployments of Train-Ticket~\cite{DBLP:journals/tse/ZhouPXSJLD21}, an open-source microservice benchmark system, with and without common resilience mechanisms (Section~\cref{sec:motivation:difference}).
We find that failures affect system performance metrics (e.g., memory usage, network throughput) such as memory usage and network throughput, but a resilient service can prevent the impact from disseminating to user-aware metrics such as response latency and MTTR.
Based on the observation, our insight is that we can measure microservice resilience by comparing the degree of degradation dissemination from system performance metrics to user-aware metrics.
\textit{If the degradation cannot disseminate from system performance metrics to user-aware metrics, the resilience is high. Otherwise, the resilience is low.}

Motivated by this insight, we present \textbf{\frmwk}, the first versatile resilience profiling framework for microservice systems.
\frmwk consists of three phases, i.e., \textit{failure execution}, \textit{dissemination-based metric lattice search}, and \textit{resilience indexing}.
\textit{Failure execution} comprises two phases: \textit{failure injection} and \textit{failure clearance}.
Given a specified failure and a predefined \textit{load generator}, \frmwk collects the service's monitoring metrics in the normal and faulty period.
For the \textit{dissemination-based metric lattice search}, we propose a dissemination-based algorithm that ranks all the monitoring metrics according to their contributions to the overall service degradation.
We construct a metric lattice from the power set of the monitoring metric set.
The ranking is based on a degradation-based path search in the metric lattice.
Lastly, for \textit{resilience indexing}, we index the resilience in $(0,1)$ by how much the degradation in system performance metrics is disseminated to the user-aware metrics.

Experiments on two open-source (\TT~\cite{DBLP:journals/tse/ZhouPXSJLD21} and \SN~\cite{DBLP:conf/asplos/GanZCSRKBHRJHPH19}) and one industrial (\company) microservice system demonstrate the effectiveness of \frmwk.
We inject failures into all systems and compare the performance of resilience profiling under \frmwk and several baselines.
The experimental results demonstrate that our proposed method accurately quantifies the system resilience and outperforms the baselines.
Specifically, in terms of cross-entropy, \frmwk achieves the best performance of 0.3246 on the \TT benchmark, 0.3766 on the \SN benchmark, and 0.2977 on the industrial benchmark.
In terms of accuracy, \frmwk also achieves the best performance of 0.9012, 0.8611, and 0.8929 on the \TT, \SN, and industrial benchmarks.
Furthermore, we showcase the successful usage of \frmwk in the production cloud system of \company.
We make the code and dataset publicly available\footnote{\textcolor{blue}{\url{https://github.com/yttty/MicroRes}}}.

The contributions of this work are highlighted as follows:
\begin{itemize}[leftmargin=10pt, topsep=2pt]
    \item We identify the labor-intensity and flexibility issues for the current rule-based practice for resilience profiling. Then we conduct the first investigation on how degradation disseminate from system performance metrics to user-aware metrics in resilient and non-resilient microservice systems, which demonstrates the viability of versatile resilience measuring.
    \item We propose \frmwk, the first versatile resilience profiling framework that can automatically index the resilience of a microservice system to different failures. \frmwk measures the dissemination of degradation from system performance metrics to user-aware metrics. The higher the dissemination, the lower the resilience.
    \item Evaluation of \frmwk on two open-source and one industrial microservice systems indicates its effectiveness and efficiency. The industrial case study also confirms the practical usage of \frmwk.
\end{itemize}

\section{Background}
\label{sec:background}
This section first briefly describes the metrics of a microservice system.
Then we present the necessity and procedure for resilience profiling that underpins our approach.

\subsection{Metrics of Microservices}
\label{sec:background:metrics}

In the contemporary landscape, large-scale online services such as Netflix and Twitter adopt microservices~\cite{microservice-architecture} to achieve scalability, robustness, and agility. This architectural approach involves breaking down a monolithic online service into fine-grained components known as microservices~\cite{DBLP:books/lib/Newman15}. These microservices are highly decoupled and can be numerous in a system; for instance, Netflix employs hundreds to thousands of microservices~\cite{Gremlin}. Virtualized infrastructure, like virtual machines and containers, is commonly used for deploying microservices.
To facilitate service decoupling and orchestration, additional components like API gateways, service registries, and databases are employed.

As a result of this complex setup, microservice systems generate extensive and diverse monitoring metrics~\cite{theory-of-monitoring}, which can vary based on the system's architecture and implementation. Broadly, these monitoring metrics fall into two categories: \textit{system performance metrics} and \textit{user-aware metrics}.

\textit{System performance metrics} directly reflect the runtime status of microservices and the underlying orchestration system.
Microservice orchestration platforms like Kubernetes~\cite{kubernetes} use multi-level isolation to manage containers in isolated pods on nodes, either virtual or physical machines. Components for network management, proxy, and task scheduling are also monitored for various system performance metrics, including CPU and memory usage, network throughput, disk I/O, TCP connections, etc., at both the infrastructure and container levels.
As any failure of these components may possibly result in the degradation of service, all the pods, nodes, and other components are monitored, producing various system performance metrics, e.g., CPU and memory usage, network throughput, network transmit and receive rate, disk I/O speed and error rate, number of TCP connections, etc.
The system performance metrics are collected at different virtualization levels, i.e., the infrastructure level (machines) and the container level (microservices).

\textit{User-aware metrics}, in addition, reflect the quality of service in a specific time period from the users' aspect.
User-aware metrics, such as response latency, error rate, throughput, mean time to recovery, and availability rate, are also crucial system indicators.
Different online services value different user-aware metrics.
For example, availability and error rate are common performance attributes of transactional services, while video streaming services are usually based on throughput.

\subsection{Microservice Resilience Testing}
\label{sec:background:resilience}

Resilience in a microservice system pertains to its capacity to sustain service performance at an acceptable level and efficiently recover from failures that lead to service degradation~\cite{resiliency-definition,DBLP:journals/ijseke/YinD21}. The construction of robust online services becomes imperative, given the inevitability of faults and failures~\cite{gray-failure,DBLP:conf/hotos/LiuLMN19}. The ability to withstand unexpected failures is crucial for minimizing downtime, upholding service quality, and fulfilling service-level agreements, which is crucial for user experience.

Resilience testing~\cite{ibm-resilience-test} is a primary method for ensuring software resilience, demanding that all new or updated microservices undergo these tests to validate the resilience of online services.
Unlike functional correctness tests~\cite{ammann2016introduction}, which focus on core application functions and data integrity, resilience tests deliberately introduce failures into the system under stress or chaotic conditions to assess how the microservice system performs~\cite{Gremlin}. 
Test engineers then use the observed flaws to refine the architectural design. The passing criterion involves the online service continuing to deliver acceptable performance despite the induced failures.
In real-world scenarios, industrial practitioners also employ chaos engineering~\cite{basiri2016chaos,DBLP:conf/issre/BlohowiakBHR16} to assess software resilience within production environments with live traffic. The resilience testing procedure encompasses \textit{failure injection} and \textit{test results determination}~\cite{techtarget-blog}.

As an example, testing the resilience of an online service when facing high network packet loss involves several steps.
First, test engineers introduce network packet loss failures by utilizing appropriate tools.
Next, they collect relevant monitoring metrics based on the engineers' domain knowledge.
In this sample case, monitoring metrics like network transmit and receive rate and request throughput will be selected.
Once the metrics are gathered and visualized (Figure~\ref{fig:faulty_metrics}), engineers can examine the data, with the green area representing the normal period and the red area signifying the faulty period.
By comparing the duration and magnitude of the monitoring metrics, the engineers can draw conclusions about the online service's resilience to network packet loss.
If the throughput experiences a significant drop during the faulty period, it indicates that the online service has failed this resilience test.

While automation of resilience testing is feasible, it remains cumbersome, necessitating the definition of test rules.
To standardize the procedure, test engineers manually determine a set of rules for each failure type, consisting of five components: \textit{failure type}, \textit{load}, \textit{monitored metrics}, \textit{degradation profiling}, and \textit{pass criteria}.
The \textit{failure type} denotes the specific failure to inject, with the expectation that the tested service should demonstrate resilience to this failure. \textit{Load} is determined based on the maximum load the service can handle without performance issues. \textit{Monitored metrics} are selected to clearly manifest the degradation caused by the injected failure, encompassing I/O rates, throughput, mean time to recovery, latency, and other relevant metrics. \textit{Degradation profiling} quantifies the degree of degradation, often considering the duration and magnitude changes of the target metrics. \textit{Pass criteria} are then established to determine the resilience test result, involving the analysis of monitoring metrics under normal and faulty conditions, leading to a \passfail conclusion. These criteria should be based on the anticipated service quality.
For each failure type, test engineers need to adhere to the outlined procedure to conduct resilience tests.

\vspace{-0.1in}
\section{Motivation}
\label{sec:motivation}

In the current industrial practice, test engineers conduct resilience profiling by manual configuration of rules.
This section first points out that the current resilience testing practice suffers from labor-intensity and flexibility issues due to the decoupled, dynamic, and specialized nature of microservices (\cref{sec:motivation:issues}).
To keep up with the fast-evolving microservices, automated and versatile resilience profiling is desired.
To explore the opportunity to automate resilience profiling, we compare the differences of failures' manifestations in the monitoring metrics between resilient and non-resilient deployments of an open-source benchmark microservice system (\cref{sec:motivation:difference}).
Our insight is that versatile resilience profiling can be automated by quantifying the degradation disseminated from the system performance metrics to the user-aware metrics.

\subsection{Issues of Current Practice}
\label{sec:motivation:issues}

\begin{table*}[t]
  \centering
  \scriptsize
  \caption{Failures and the corresponding degradation with and without the resilience mechanisms mentioned in \cref{sec:motivation:difference}}
  \label{tab:failures}

  \begin{tabular}{|l|l|p{3.4cm}|p{5.1cm}|p{4.5cm}|}
    \hline
    \textbf{Virtualization Level}                            & \textbf{Type}                      & \textbf{Failure}                   & \textbf{Degradation w/o resilience mechanisms}       & \textbf{Degradation w/ resilience mechanisms} \\ \hline
    \multirow{17}{*}{\textit{Infrastructure}} & \textit{CPU}                       & CPU overload                       & High physical CPU usage, slow response speed         & Decreased but acceptable response speed       \\ \cline{2-5}
                                              & \textit{Memory}                    & Memory overload                    & High physical memory usage, slow response speed      & Decreased but acceptable response speed       \\ \cline{2-5}
                                              & \multirow{5}{*}{\textit{Storage}}  & Disk partition full                & Unable to read/write, internal error (500)           & Normal response                             \\ \cline{3-5}
                                              &                                    & High disk I/O throughput           & High physical I/O throughput                         & Normal response                             \\ \cline{3-5}
                                              &                                    & High disk I/O latency              & Slow I/O                                             & Normal response                             \\ \cline{3-5}
                                              &                                    & High disk I/O error                & Slow and erroneous I/O                               & Normal response                             \\ \cline{3-5}
                                              &                                    & Block storage service stopped      & I/O rate drop to zero, internal error (500)          & Normal response                             \\ \cline{2-5}
                                              & \multirow{6}{*}{\textit{Network}}  & High HTTP packet loss rate         & High retransmission rate                             & Normal response                             \\ \cline{3-5}
                                              &                                    & High HTTP request latency          & High connection latency, slow response               & Return to normal response speed shortly       \\ \cline{3-5}
                                              &                                    & TCP disconnection                  & Connection error, disconnected                       & Return to normal response speed shortly       \\ \cline{3-5}
                                              &                                    & Port in use                        & Connection initialization error                      & (same as left)                                \\ \cline{3-5}
                                              &                                    & NIC down                           & Connection error, unreachable network                & (same as left)                                \\ \cline{3-5}
                                              &                                    & Running out of network connections & Unable to create new connections                     & Normal response                             \\ \cline{2-5}
                                              & \textit{Process}                   & Critical process killed            & Unresponsive process, existing connection down       & Normal response after some time             \\ \cline{2-5}
                                              & \multirow{3}{*}{\textit{Machine}}  & Unplaned reboot                    & Machine offline                                      & Normal response after some time             \\ \cline{3-5}
                                              &                                    & Power outage                       & Machine offline                                      & Normal response after some time             \\ \cline{3-5}
                                              &                                    & System time shift                  & Process error                                        & Automaitc time correction                     \\ \hline
    \multirow{10}{*}{\textit{Container}}      & \textit{CPU}                       & Container CPU overload             & High container CPU usage, slow response speed        & Decreased but acceptable response speed       \\ \cline{2-5}
                                              & \textit{Memory}                    & Container memory overload          & High container memory usage, slow response speed     & Decreased but acceptable response speed       \\ \cline{2-5}
                                              & \multirow{5}{*}{\textit{Network}}  & Container TCP disconnection        & Connection error within container                    & Return to normal response speed shortly       \\ \cline{3-5}
                                              &                                    & Unreachable network                & Network unreachable error in container               & Return to normal response speed shortly       \\ \cline{3-5}
                                              &                                    & Container port in use              & Connection initialization error                      & (same as left)                                \\ \cline{3-5}
                                              &                                    & Container network packet loss      & High retransmission rate                             & Return to normal response speed shortly       \\ \cline{3-5}
                                              &                                    & Container virtual NIC down         & Connection error                                     & Return to normal response speed shortly       \\ \cline{2-5}
                                              & \textit{Storage}                   & Container disk full                & Unable to read/write, internal error (500)           & Normal response after some time             \\ \cline{2-5}
                                              & \multirow{2}{*}{\textit{Instance}} & Container instance killed          & Instance offline, unresponsive microservice endpoint & Normal response after some time             \\ \cline{3-5}
                                              &                                    & Container instance suspended       & Instance offline, unresponsive microservice endpoint & Normal response after some time             \\ \hline
  \end{tabular}
\end{table*}

Currently, test engineers manually set resilience test rules for each service and each failure type.
Setting the rules heavily depends on human expertise.
As demonstrated below, such a practice suffers from labor-intensity and flexibility issues, especially when evaluating the resilience of an online service composed of multiple fast-evolving microservices.

\subsubsection{\textbf{Labor-intensity Issue}}
Cloud providers are increasingly becoming worried of relying on manual labor and expertise for resilience profiling. This is partly because the process of creating rules is time-consuming and labor-intensive. The problem of labor intensity is especially pronounced in microservices, primarily due to two specific reasons.

\begin{itemize}[leftmargin=10pt, topsep=2pt]
    \item[1)] \textit{Decoupled, massive components.} Since microservice systems are highly decoupled, the number of microservices is very large. Due to the complex dependency~\cite{DBLP:conf/nsdi/ZhaiCPBTSZ20, AID} and system architecture~\cite{k8s-doc-arch}, the number of failures increases exponentially with the number of microservices in the system. Making proper rules under such massiveness is really challenging.
    \item[2)] \textit{Dynamics.} Microservices encourage seamless updates and flexible deployment of services~\cite{berkeley-view-cloud}, so the failure rule sets should be updated accordingly, incurring lots of burden on test engineers.
\end{itemize}

Our investigation into a cloud service provider, \company, reveals that each service has around 26 microservices on average, with the largest having over 190 microservices. Each microservice generates over 40 metrics, resulting in approximately 1040 monitoring metrics per cloud service. Despite the possibility of automating the analysis of monitoring metrics, manual resilience rule definition and updating remain labor-intensive, taking about two person-months per cloud service.
As a result, the manual identification of resilience test rules is not too time-consuming and labor-intensive for large-scale microservices.

\subsubsection{\textbf{Flexibility Issue}}
Fixed resilience test rules cannot fit different microservice systems, as well as microservices with various refined resilience mechanisms, lacking the desired adaption to different systems.
We attribute this flexibility issue to two reasons.

\begin{itemize}[leftmargin=10pt, topsep=2pt]
    \item[1)] \textit{Diversity exists in micoservices and their failures.} Microservices are specialized and may fail in different ways~\cite{DBLP:conf/hotos/LiuLMN19}, so the manifestations of failures are also manifold. The current practice requires per-system and per-fault re-configurations on test rules.
    \item[2)] \textit{Resilience sometimes is not an either-or thing.} The boundary between "resilient" and "non-resilient" in certain cases is less absolute due to the presence of refined resilience mechanisms.
        Fixed resilience test rules with binary \passfail results may not adequately capture the subtle differences in an online service's resilience for two main reasons.
        First, the impact of failures in a microservice system is diverse, as the decoupled architecture~\cite{k8s-doc-arch} often leads to partial failures of microservices~\cite{aws-challenges-with-distributed-systems}.
        Second, online services adopting the microservice architecture commonly employ multiple ways for fault tolerance, e.g., multiple replications and active traffic control~\cite{k8s-doc-disruption}.
        With these fault tolerance mechanisms, the online service can be in a gray-failure status~\cite{gray-failure}.
\end{itemize}

For example, suppose we conduct resilience tests on an online service.
The passing criteria require the mean time to recovery to be 5 minutes, which means the microservice should recover to the normal status in 5 minutes after the failure injection.
Given the throughput of a microservice's two versions A and B under the same failure, the only difference is that version A takes 5 minutes to recover while version B only takes 2 minutes.
Version B has higher resilience than A.
However, both versions {\small \texttt{PASS}} the resilience test and we cannot explicitly know which one is more resilient.
Thus, fixed rules cannot reflect the subtle difference in resilience.

Compared with traditional monolithic applications, the metric analysis for a microservice system becomes more complex because
(1) the decoupled and specialized nature of microservices makes the number of monitoring metrics explode, and
(2) the mutual influence between monitoring metrics becomes exquisite~\cite{DBLP:conf/asplos/GanZHCHPD19,DBLP:conf/asplos/0002LD0D21}.

To sum up, the current manual test rule configuration suffers from the labor-intensity issue due to the decoupled and dynamic attributes of microservices.
The impact of failures is diverse. The fixed test rules cannot adapt to different microservice systems and cannot depict the subtle difference in an online service's resilience, which results in the flexibility issue.
Thus, it is necessary to design a framework for resilience testing that can automatically adapt to different failures without defining the rules manually.

\vspace{-0.1in}
\subsection{Investigation on Failures' Impact}
\label{sec:motivation:difference}

Microservice resilience is frequently compromised due to ubiquitous failures~\cite{DBLP:conf/nsdi/ZhaiCPBTSZ20, DBLP:journals/ijseke/YinD21}, including the inherent bugs~\cite{DBLP:journals/tse/ZhouPXSJLD21, DBLP:conf/sigsoft/Zhou0X0JLXH19}, unstable message passing~\cite{DBLP:conf/issre/JagadeesanM20}, and unreliable cloud infrastructure~\cite{aws-past-event, DBLP:conf/hotos/LiuLMN19}. 
Even routine operations, such as software upgrades and configuration file changes, can lead to significant service disruption~\cite{DBLP:conf/cloud/GunawiHSLSAE16}.

To identify microservice resilience failures, we analyzed incident reports from 2020 to 2022 at \company. Two senior Ph.D. students, familiar with the cloud computing system, classified each failure by level (infrastructure or container) and type (e.g., memory, network, machine).
We collected failures that occurred one or more times and were related to service resilience, with input from an experienced cloud system architect. The analysis yielded 27 relevant failures, categorized by virtualization level and type of failed resource. Software bugs were excluded as they are typically detected through functional testing. Table~\ref{tab:failures} lists these failures.

To comprehend the impact of failures, we conduct an empirical study on two different deployments of the \TT open-source microservice benchmark system~\cite{DBLP:journals/tse/ZhouPXSJLD21}. One deployment is configured with common resilience mechanisms (load balancing and two replications for each microservice), while the other lacks these mechanisms. The study takes place on a Kubernetes cluster with 128 GB memory and 24 CPU cores, and monitoring metrics are collected and visualized using cAdvisor~\cite{cadvisor} and Prometheus~\cite{prometheus}.

We inject the failures listed in Table~\ref{tab:failures} into the Kubernetes cluster using ChaosBlade~\cite{chaosblade} and record and analyze the system's response. Finally, we compare the impacts with and without the common resilience mechanisms.

The impact of the injected failures becomes evident through service degradation.
This degradation is quantified by measuring how much the service's performance deviates from the benchmark~\cite{DBLP:journals/ijseke/YinD21}. We herein use the service's average performance without injected failures as the benchmark.
The service degradation is determined by comparing the performance during the normal period with that during the fault-injection period. Table~\ref{tab:failures} contains the failure manifestations without applying the described resilience mechanisms in the penultimate column, while the last column shows the failure manifestations with the resilience mechanisms applied. It's worth noting that the same failure may cause different degrees of degradation depending on the employed resilience mechanism.

Through a comparative analysis of the last two columns in Table~\ref{tab:failures}, we observe that failures can exhibit diverse impacts and resilient services can mitigate the impact of failures on system performance metrics, while user-aware metrics are less affected.
For instance, when there is only one container, the container CPU overload failure leads to 100\% CPU usage remaining for an extended duration and affects the end user's experience negatively.
Nevertheless, when multiple replications are employed, the impact on user-aware metrics, such as throughput, becomes less severe.
Another example is that microservices with two active replications can rapidly recover from a "container instance killed" failure. Conversely, microservices lacking such replication mechanisms will experience extended recovery times or even break down entirely.

\subsection{Our Insight}
In this paper, we define \textbf{degradation dissemination} as \textit{the process by which degradation in system performance metrics spreads or disseminates to affect user-aware metrics in a microservice system.}
When a failure causes degradation in the system performance metrics, it can have an impact on user-aware metrics, leading to less resilient services.

As evidenced by our empirical study on failures' impact, we suggest that versatile and labor-efficient resilience profiling can be achieved by analyzing the dissemination of degradation from system performance metrics to user-aware metrics. When the degradation of user-aware metrics mirrors system performance metrics, the failure's impact spreads from the system to the user-aware level, resulting in less resilient services. Conversely, lower dissemination of degradation implies higher microservice system resilience. This finding highlights the possibility of creating a versatile framework for assessing a microservice system's resilience to different failures, eliminating the need for manually defining resilience test rules.

\begin{mybox}
    \small
    \textbf{Insight}: Microservices exhibit diverse failure patterns, resulting in various impacts on metrics. The primary consequence of these failures is service degradation. Higher resilience is associated with limited dissemination of degradation from system performance metrics to user-aware metrics.
\end{mybox}
\vspace{-0.05in}

\section{Methodology}
\label{sec:approach}

\begin{figure*}[t]
    \centering
    \includegraphics[width=2\columnwidth]{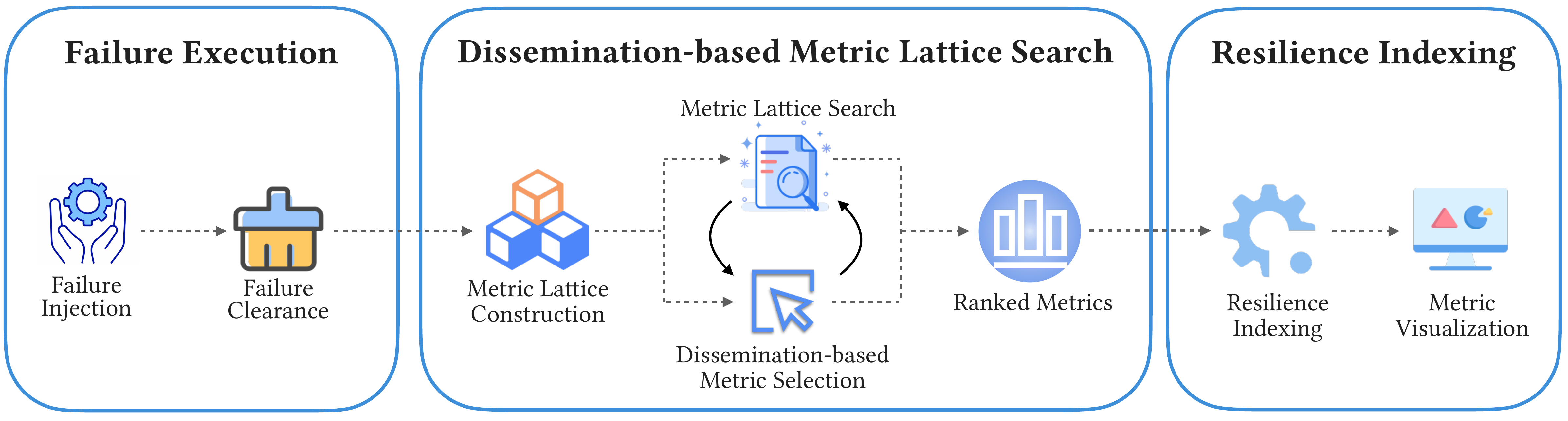}
    \caption{Overall framework of \frmwk.}
    \label{fig:overview}
\end{figure*}

We propose \frmwk, a versatile microservice resilience profiling framework via degradation dissemination indexing.
Figure~\ref{fig:overview} illustrates the overall workflow of \frmwk.
It consists of three phases, i.e., \textit{failure execution}, \textit{dissemination-based metric lattice search}, and \textit{resilience indexing}.
The \textit{failure execution} is composed of failure injection and failure clearance.
Given a specified failure and a predefined load generator, \frmwk collects the to-be-tested service's monitoring metrics in the normal and faulty period.
For the \textit{dissemination-based metric lattice search}, we propose a dissemination-based metric selection algorithm.
We organize all possible metric subsets of the monitoring metrics as a huge lattice.
Then \frmwk searches the lattice while reducing the dimension by gradually selecting and removing the metric that contributes most to the overall degradation.
In this way, the search path naturally forms a ranked list of monitoring metrics along with their contribution to the overall degradation.
Lastly, for \textit{resilience indexing}, we calculate the resilience index by how much the degradation disseminates from system performance metrics to user-aware metrics.

\frmwk measures the performance loss (i.e., the degree of service degradation) by comparing the metrics' difference between normal and faulty periods.
It quantifies the degradation dissemination by ranking the monitoring metrics' contribution to the overall degradation.
In short, if system performance metrics contribute more to overall degradation than user-aware metrics, degradation dissemination is less, indicating higher resilience.

Such design addresses the labor-intensity and flexibility issues.
First, \frmwk automatically produces resilience indices by measuring the degradation dissemination from system performance metrics to user-aware metrics, significantly alleviating human labor and saving time.
Second, as \frmwk uses ranking, it is agnostic to the system architecture or adopted resilience mechanisms, allowing for flexible adoption to different microservice systems without system-dependent or fault-specific configurations.

\subsection{Failure Execution}
\label{sec:approach:failure}

The \textit{failure execution} consists of the \textit{failure injection} and the \textit{failure clearance} phases.
First, a test engineer needs to provide a \textit{load generator} to the online service being tested.
The load generator should mimic real-world requests from users.
Second, the test engineer selects a list of failures to test.
The failure can be injected at the infrastructure level or at the container level.
Then, \frmwk automatically generates a failure injection pipeline.
For each failure, \frmwk injects the failure, clears the failure, and collects the service's monitoring metrics in the meantime.
The duration of failure injection and failure clearance are the same for each failure.

During the two phases, \frmwk collects two types of metrics, i.e., user-aware metrics and system performance metrics.
Suppose $\mathcal{B}$ is the user-aware metrics set and $\mathcal{P}$ is the system performance metrics set in the system.
We denote the set of all the user-aware metrics and system performance metrics as $\mathcal{M} = \mathcal{B} \cup \mathcal{P}$.
Suppose $card(\mathcal{M}) = M$, we can index all the monitoring metrics from $m_1$ to $m_M$.
In other words, $\mathcal{M} = \{ m_1, m_2, \cdots, m_M \} $.
Thus, for any $i \in [1,M]$, either $m_i \in \mathcal{B}$ or $m_i \in \mathcal{P}$.
We denote the monitoring metrics during the failure injection period as $\mathcal{M}^f = \{ m^f_1, m^f_2, \cdots, m^f_M \}$.
For each $i$, $m_{i}^f$ is a univariate time series denoting the monitoring metrics during the failure injection (faulty) period.
Likewise, we denote the monitoring metrics during the failure clearance (normal) period as $\mathcal{M}^n = \{ m^n_1, m^n_2, \cdots, m^n_M \}$.
Also, for each $i$, $m_{i}^n$ is a univariate time series denoting the monitoring metrics during the failure clearance (normal) period.
We ensure that $length(m_i^f) = length(m_i^n) = T$.

\subsection{Dissemination-based Metric Lattice Search}
\label{sec:approach:demeter}

The dissemination-based metric lattice search aims at comparing and ranking the contribution of different monitoring metrics to the overall service degradation caused by the failure.
Algorithm~\ref{algo:demeter_overall} shows the procedure for dissemination-based metric lattice search.
We introduce the dissemination-based metric lattice search from the following three aspects, i.e., \textit{metric lattice construction}, \textit{dissemination-based metric selection}, and \textit{metric lattice search}.

\begin{algorithm}[htbp]
    \small
    \caption{Dissemination-based Metric Lattice Search}
    \label{algo:demeter_overall}
    \LinesNumbered
    \KwIn{The monitoring metrics $\mathcal{M} = \{ m_1, m_2, \cdots, m_M \} $; The monitoring metrics during the failure injection period $\mathcal{M}^f = \{ m^f_1, m^f_2, \cdots, m^f_M \}$; The monitoring metrics during the failure clearance period $\mathcal{M}^n = \{ m^n_1, m^n_2, \cdots, m^n_M \}$}
    \KwOut{An ranked list of metrics $\hat{\mathcal{M}}$}
    Construct the metric lattice (Section~\cref{sec:approach:demeter:lattice_construct}) \\ 
    $\mathcal{L} = EmptyList() $ \\
    $M = \mathcal{M} $ \\
    \While(// Metric Lattice Search){$M \neq \emptyset$}{
      $cmax,m_{imax} = \mathtt{MetricSelection}(M)$ \\
      $\mathcal{L}.append((cmax,m_{imax}))$ \\
      $M = M - \{ m_{imax} \}$ \\
    }
    \Return{$\mathcal{L}$}
\end{algorithm}

\subsubsection{Metric Lattice Construction}
\label{sec:approach:demeter:lattice_construct}

Formally, a lattice is a partially ordered set in which each pair of elements has a least upper bound and a greatest lower bound.
Inspired by the frequent itemset mining algorithm~\cite{morishita2000transversing,han2011data}, we construct a lattice from the power set (i.e., the set of all subsets) of all the available monitoring metrics (denoted as $\mathcal{M}$).
Let each subset of $\mathcal{M}$ be a node in the metric lattice $\mathcal{L}$.
We define the order between any two nodes of the lattice as the subset-superset relation.
Formally, suppose we have $a, b \subseteq \mathcal{M}$ and $a \neq b$, then $a \subset b (\subseteq \mathcal{M})$ (in the monitoring metric set) indicates $a \leq b$ ($b \rightarrow a$ in the metric lattice).
Given the definition, for any $a$ and $b$, the least upper bound is $\mathcal{M}$. The greatest lower bound is $\emptyset$.
Hence, the correctness of the generated lattice is theoretically guaranteed.
The metric lattice will be searched starting from the node $\mathcal{M}$ in later steps.

\begin{figure}[htbp]
    \centering
    \includegraphics[width=\columnwidth]{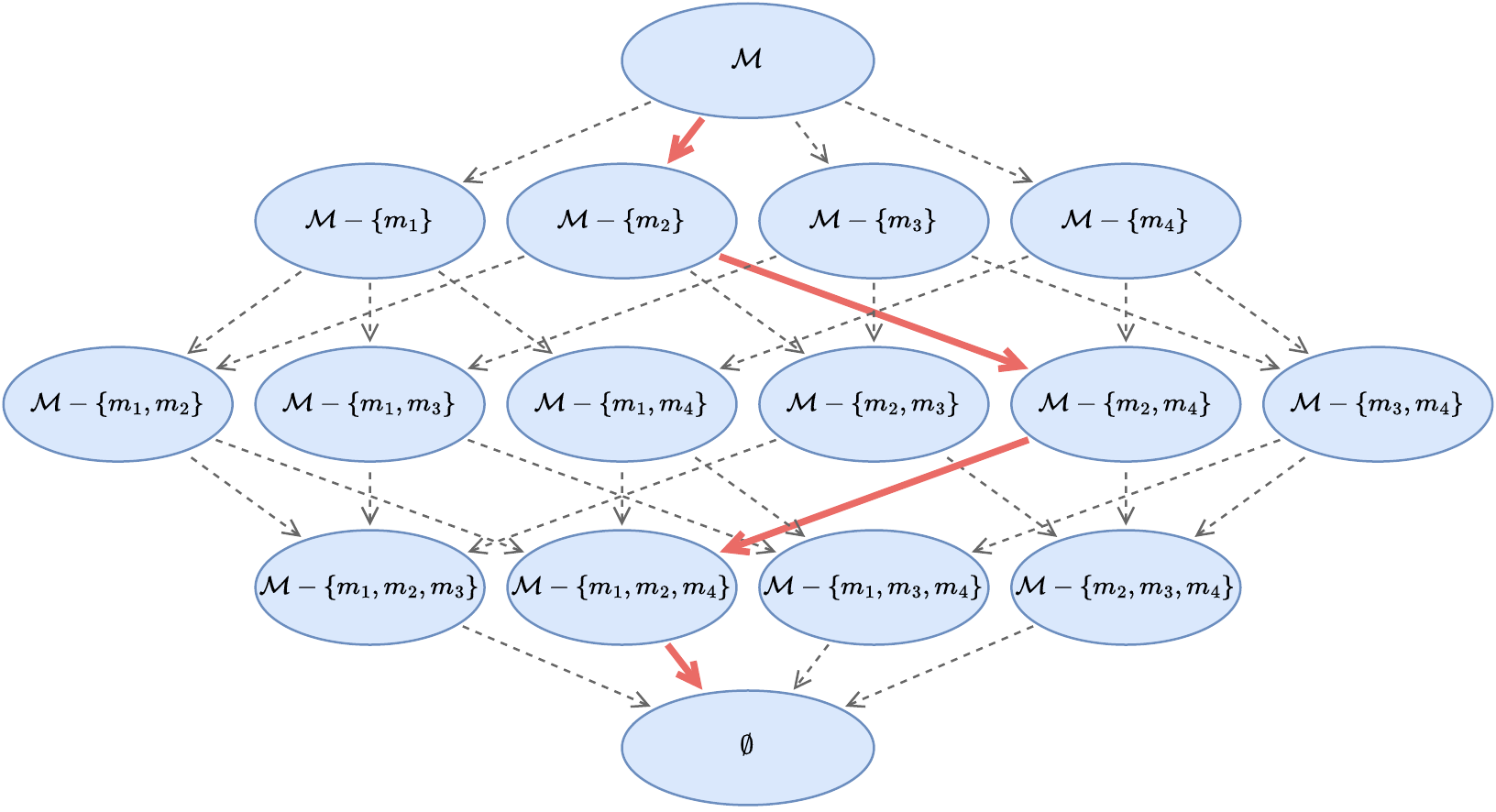}
    \vspace{-0.05in}
    \caption{An example metric lattice constructed from $\mathcal{M} = \{m_1,\cdots,m_4 \}$. We set the number of monitoring metrics as a small value, $4$, for a clear illustration. The path of all solid red edges forms a ranked list.}
    \label{fig:lattice}
    \vspace{-0.1in}
\end{figure}

Figure~\ref{fig:lattice} illustrates an example metric lattice constructed from $\mathcal{M} = \{m_1,\cdots,m_4 \}$.
Each directed edge indicates a subset-superset relation, pointing from the metric superset to the metric subset.
Note that we set the number of monitoring metrics as a small value, $4$, for a clear illustration.

\subsubsection{Dissemination-based Metric Selection}
\label{sec:approach:demeter:degradation}

As mentioned in Section~\cref{sec:motivation:difference}, service degradation is the primary manifestation of the failures' impact.
We propose to measure the service degradation via the fluctuation of system performance metrics and user-aware metrics.
If the degradation of system performance metrics cannot disseminate to the degradation of user-aware metrics, resilience is higher.
Otherwise, the resilience is lower.
Therefore, the key is to select the monitoring metric that contributes most to the overall service degradation among all the monitoring metrics.

\begin{algorithm}[htbp]
    \small
    \caption{Dissemination-based Metric Selection}
    \label{algo:demeter_degradation}
    \LinesNumbered
    \KwIn{The monitoring metric subset $\mathcal{M}'$; The monitoring metrics during the failure injection period $\mathcal{M}'^f$; The monitoring metrics during the failure clearance period $\mathcal{M}'^n$}
    \KwOut{The metric $m_i \in \mathcal{M}'$ where $m_i$ contribute most to the overall service degradation}

    \SetKwFunction{FDeg}{MetricSelection}
    \SetKwProg{Fn}{Function}{:}{End}
    \Fn{\FDeg{$\mathcal{M}'$, $\mathcal{M}'^f$, $\mathcal{M}'^n$}}{
      $T = $ length of the monitoring metrics \\
      $\mathbf{D} = []$ \\
      \For{$m_i \in \mathcal{M}' $}{
        // Compute the performance difference of each individual metric \\
        \For{$t = 1 \ldots T $}{
          $\delta_i(t) = |m_i^f(t) - m_i^n(t)|$ \label{algo:demeter_degradation:corr}
        }
        $\hat{\delta_{i}} = \delta_{i} - \bar{\delta_{i}}$ // Normalize $\delta_{i}$ \\
        $\mathbf{D} = [\mathbf{D}; \hat{\delta_{i}}]$ // Concatenate the normalized performance difference \label{algo:demeter_degradation:concat} \\
      }
      $\delta_{PC1} = \mathtt{PCA}(\mathbf{D}, dim=1) $ // Reduce to one dimension via Principal Component Analysis \label{algo:demeter_degradation:pca} \\
      // Select the metric that contributes most to the performance difference \\
      \For{$\hat{\delta_{i}} \in \mathbf{D} $}{
        $c_i = \mathtt{Contribution}(\delta_{PC1}, \hat{\delta_{i}})$  \label{algo:demeter_degradation:contri} \\
      }
      $cmax = \max (c_i)$ \\
      $imax = {\arg\max}_i (c_i)$ \label{algo:demeter_degradation:pca_end}  \\
      \KwRet $cmax$, $m_{imax}$ \\
    }
\end{algorithm}

Algorithm~\ref{algo:demeter_degradation} shows how to select the metric that contributes most to the overall service degradation.
Expressly, given a subset of the entire monitoring metrics set $\mathcal{M}' \subseteq \mathcal{M}$ and the metrics during the faulty and normal period $\mathcal{M}'^f$ and $\mathcal{M}'^n$.
We first compute the performance difference $\delta_i$ of each monitoring metric $m_i$ (Line \ref{algo:demeter_degradation:corr}).
The computation involves determining the absolute difference for each specific metric during the failure injection period, paired with the metrics during the failure clearance period. This absolute difference serves as a measure of the influence of injected failures on each metric.
All metrics' performance difference naturally forms a performance difference matrix $\mathbf{D}$ (Line \ref{algo:demeter_degradation:concat}).
Subsequently, we identify the metric that has the most significant impact on the performance difference, as outlined in Lines \ref{algo:demeter_degradation:pca} to \ref{algo:demeter_degradation:pca_end}.

We apply Principal Component Analysis (PCA)~\cite{pca, han2011data} on the performance difference matrix $\mathbf{D}$ to reduce $\mathbf{D}$ to 1 dimension.
PCA is a statistical technique for simplifying and understanding complex data by reducing its dimensionality while preserving most of its variability.
In our case, we have a bunch of metrics' performance differences in a high-dimensional space. Each dimension represents a different attribute of degradation, e.g., disk I/O, network, response latency, etc.
We use PCA to find a new dimension, called the principal component, that captures the most important performance difference.
The principal component, of length $T$ (Line \ref{algo:demeter_degradation:pca}), denoted as $\delta_{PC1}$, is a linear combination of the original metrics principal components.
Let $\delta_{PC1}$ represent the overall service degradation caused by the injected failure.
We compute the contribution of each metric to the overall degradation via a contribution measure \texttt{Contribution()} (Line \ref{algo:demeter_degradation:contri}).
The higher the similarity between $\delta_{PC1}$ and $\delta_{i}$, the larger the \texttt{Contribution()} outputs.
\texttt{Contribution()} can be a correlation coefficient (e.g., Pearson correlation coefficient) or any other distance measure (e.g., Euclidean distance or dynamic time warping distance) deemed appropriate.
We discuss the selection of \texttt{Contribution()} in Section~\cref{sec:experiment:rq4}.
In the end, the function returns the metric $m_{imax}$ that contributes most to the overall service degradation, along with its contribution $cmax$.
$m_{imax}$ will guide the metric lattice search, and $cmax$ will be used to calculate resilience in Section~\cref{sec:approach:indexing}.

\begin{table*}[ht]
\centering
\caption{Effectiveness Comparison (RQ1) and Ablation Study (RQ2) of \frmwk}
\vspace{-0.1in}
\label{tab:rq3}
\begin{adjustbox}{max width=2.1\columnwidth}
\begin{tabular}{c*{17}{c}}
\toprule
\multirow{2}*{\centering \textbf{Category}} & \multirow{2}*{\centering \textbf{Method}} & \multicolumn{5}{c}{\textit{\TT}} & \multicolumn{5}{c}{\textit{\SN}} & \multicolumn{5}{c}{\textit{\IND}} \\
\cmidrule(lr){3-7}\cmidrule(lr){8-12}\cmidrule(lr){13-17}
& & \textit{CE} & \textit{MAE} & \textit{RMSE} & \textit{Acc} & \textit{F1} & \textit{CE} & \textit{MAE} & \textit{RMSE} & \textit{Acc} & \textit{F1} & \textit{CE} & \textit{MAE} & \textit{RMSE} & \textit{Acc} & \textit{F1} \\
\cmidrule{1-17}\morecmidrules\cmidrule{1-17}
\multirow{3}*{\centering \textbf{RQ1}} & SVC 
& 0.8830 & 0.3497 & 0.5267 & 0.5802 & 0.7018
& 1.2608 & 0.3908 & 0.5657 & 0.5278 & 0.6383
& 0.6743 & 0.3786 & 0.4627 & 0.6786 & 0.7273 \\
& RF
& 0.9399 & 0.3507 & 0.5277 & 0.5802 & 0.7018
& 0.6708 & 0.2358 & 0.4063 & 0.5833 & 0.6809
& 0.7477 & 0.4012 & 0.4865 & 0.5000 & 0.5882 \\
& ET
& 0.8163 & 0.2999 & 0.4771 & 0.5926 & 0.7227
& 0.9160 & 0.3135 & 0.4927 & 0.6111 & 0.6818
& 0.5340 & 0.3100 & 0.3814 & 0.5714 & 0.6842 \\
\midrule
\multirow{3}*{\centering \textbf{RQ2}} & \frmwk-euc
& 0.4464 & 0.1868 & 0.3384 & 0.6543 & 0.7846
& 0.7199 & 0.2861 & 0.4640 & 0.6389 & 0.7451
& 0.4409 & 0.3036 & 0.3729 & 0.6071 & 0.7027 \\
& \frmwk-corr
& 0.3629 & 0.1730 & 0.3174 & 0.6914 & 0.8092
& 0.5969 & 0.2201 & 0.3865 & 0.6111 & 0.7407
& 0.4049 & 0.2882 & 0.3516 & 0.5714 & 0.6842 \\
& \frmwk-cid
& 0.3725 & 0.1645 & 0.3037 & 0.8148 & 0.8966
& 0.5154 & 0.1851 & 0.3326 & 0.8333 & 0.9091
& 0.3855 & 0.2737 & 0.3304 & 0.8571 & 0.9130 \\
\midrule
& \textbf{\frmwk} & \textbf{0.3246} & \textbf{0.1618} & \textbf{0.2993} & \textbf{0.9012} & \textbf{0.9481}
& \textbf{0.3766} & \textbf{0.1814} & \textbf{0.3382} & \textbf{0.8611} & \textbf{0.9231}
& \textbf{0.2977} & \textbf{0.2436} & \textbf{0.2812} & \textbf{0.8929} & \textbf{0.9362} \\
\bottomrule
\end{tabular}
\end{adjustbox}
\vspace{-0.05in}
\end{table*}

\subsubsection{Metric Lattice Search}
\label{sec:approach:demeter:search}

The metric lattice search is straightforward with the dissemination-based metric selection.
As shown in Algorithm~\ref{algo:demeter_overall}, the search starts from the node of the entire metric set $\mathcal{M}$.
At each node $\mathcal{M}' \in \mathcal{M}$, we select the metric $m_{imax}$ that contributes most to the service degradation on the metric set $\mathcal{M}'$.
We then eliminate the monitoring metric $m_{imax}$ from $\mathcal{M}'$ and proceed to the next node until all the monitoring metrics are eliminated.
The path from $\mathcal{M}$ to $\emptyset$ naturally forms an ordered list of all the monitoring metrics $m$ and their contribution value $c$, denoted as $\mathcal{L}$.
For example, in Figure~\ref{fig:lattice}, the path of all solid red edges forms the ordered list $[m_2,m_4,m_1,m_3]$.

\subsection{Resilience Indexing}
\label{sec:approach:indexing}

Section~\cref{sec:motivation:difference} finds that resilience can be inferred from whether the degradation in system performance metrics disseminates to the degradation in user-aware metrics.
To quantify the degradation dissemination, we calculate the degradation in system performance metrics and user-aware metrics with Equation~\ref{eq:deg_p} and Equation~\ref{eq:deg_b}, respectively.
Equation \ref{eq:deg_p} and \ref{eq:deg_b} are derived from the Discounted Cumulative Gain~\cite{croft2010search}, which initially measures the quality of search engines' results from the aspect of both the order and the content relevance.

\begin{align}
    D_\mathcal{P} & = \sum_{m_i \in \mathcal{P}} \frac{c_i}{\log_2 (rank(m_i;\mathcal{L}) + 1)} \label{eq:deg_p} \\
    D_\mathcal{B} & = \sum_{m_i \in \mathcal{B}} \frac{c_i}{\log_2 (rank(m_i;\mathcal{L}) + 1)} \label{eq:deg_b}
\end{align}

In the end, we utilize the sigmoid function to map the difference between $\mathcal{B}$'s and $\mathcal{P}$'s contribution to a float value $r \in (0,1)$, as shown in Equation~\ref{eq:resilience}.
\begin{equation}
    r = \frac{1}{1+e^{D_\mathcal{B}-D_\mathcal{P}}} \label{eq:resilience}
\end{equation}
where $r$ measures the degradation dissemination from the system performance metrics to the user-aware metrics.
Larger $r$ means higher resilience.
In practice, engineers can set a resilience threshold $\tau$ to get binary \passfail results, i.e., $r > \tau \Rightarrow$ \texttt{\small PASS} and $r < \tau \Rightarrow$ \texttt{\small FAIL}.

\section{Evaluation}\label{sec:experiment}

This section evaluates \frmwk by answering the following research questions:
\begin{itemize}[leftmargin=*]
    \item \textbf{RQ1.} How effective is \frmwk in evaluating the resilience of online services?
    \item \textbf{RQ2.} How do different contribution measures affect the performance of \frmwk?
    \item \textbf{RQ3.} How efficient is \frmwk?
\end{itemize}

\subsection{Experiment Settings}
\label{sec:experiment:settings}

\subsubsection{Dataset}
To illustrate the practical effectiveness of \frmwk, we carried out experiments on two simulated datasets and one industrial dataset.
Since there is no existing dataset for resilience testing, we conducted resilience tests on two open-source microservice systems and one industrial microservice system.
We collected the monitoring metrics and manually labeled the resilience testing results to build the three datasets.
We release all datasets with the paper to facilitate future research in this field.

\begin{table}[ht]
\centering
\small
\vspace{-0.05in}
\caption{Dataset Statistics}
\vspace{-0.1in}
\label{tab:dataset_statistics}
\begin{adjustbox}{max width=\columnwidth}
\begin{tabular}{@{}cccccc@{}}
\toprule
\textbf{Dataset}  & $|\mathcal{B}|$ & $|\mathcal{P}|$ & \textit{\#Microservices} & \textit{\#Failures} & \textit{Failure Duration} \\ \midrule
\textit{\TT}  & 30    & 195     & 15     & 24    & 10 minutes  \\ \midrule
\textit{\SN}  & 50    & 325     & 25     & 10    & 5 minutes  \\ \midrule
\textit{\IND} & 2     & 12      & (Undisclosed)     & 28    & 20 minutes  \\ \bottomrule
\end{tabular}
\end{adjustbox}
\vspace{-0.05in}
\end{table}

\textit{Simulated Datasets}:
For collecting the first simulated dataset, we deploy \TT \cite{DBLP:journals/tse/ZhouPXSJLD21}, an open-source microservice system, with Kubernetes, a popular microservice orchestrator.
\TT is a web-based ticketing system with 15 microservices.
For load generation, we develop a request simulator to simulate the access of ordinary users to the ticketing system.
The simulator will log in to the system, search for tickets, order tickets, food, insurance, and make the payment.
We inject 24 failures listed in Table~\ref{tab:failures} into the benchmark microservice system with ChaosBlade. (We omit the three failures in ``Infrastructure - Machine'' as we do not have any access to the physical server.)
For each failure, the failure injection period and failure clearance period both last for 10 minutes, during which the simulator continuously sends requests to the system.
cAdvisor~\cite{cadvisor} is used to collect 13 system performance metrics.
The system performance metrics cover all major aspects of the microservice system, including CPU, file system, memory, and network.
As for the user-aware metrics, we use Jaeger, an open-source tracing framework, to trace all the API calls.
Following the existing research~\cite{AID}, we calculate the average response time and the request error rate in seconds as the user-aware metrics.

Similarly, we collected the second simulated dataset on another widely used microservice orchestrator ``docker-compose''.
Different from ``Kubernetes'', ``docker-compose'' orchestrates microservices on a single host.
The resilience of a ``docker-compose'' microservice system depends more on the microservice developer.
We use the \SN \cite{DBLP:conf/asplos/GanZCSRKBHRJHPH19} microservice system.
It includes 12 microservices for processing user requests and 13 microservices for data storage.
Its user-aware metrics include the average response latency and the request error rate.
As ``docker-compose'' employs few resilience mechanisms at the infrastructure level, we only inject 10 failures at the container level with ChaosBlade.
Each failure lasts for 5 minutes since the \SN benchmark responds faster than \TT.

\textit{Industrial Dataset}:
To illustrate the practical usage of \frmwk, we collected an industrial dataset from the production cloud system of \company.
Serving tens of millions of users worldwide, \company provides many cloud services to users, including cloud virtual machines, cloud databases, edge computing, data analytics, etc.
The data analytic service adopts the microservice architecture.
We inject 27 container-level and infrastructure-level failures into the data analytic service using the proprietary fault injection tool.
As the production system takes roughly half a minute to complete one request, we let each failure last for 20 minutes.
Limited by the production system, we collected 12 performance metrics and 2 user-aware metrics in total.
The user-aware metrics of the dataset include the latency and the error rate.

\textit{Manual labeling}:
As \frmwk is unsupervised, labels are only for evaluation.
We adopt the criteria in Section~\cref{sec:motivation} for resilience, i.e., whether the degradation in system performance metrics disseminates to the degradation in user-aware metrics.
Following the existing work~\cite{AID}, we adopt binary \passfail labels since it is easier for annotators to reach an agreement.
For the industrial dataset, test engineers from \company investigate the monitoring data and annotate \passfail labels according to the criteria and their expertise.
For the simulated datasets, industrial engineers were not available to develop all the ground truth.
Thus, we invited two senior Ph.D. students to inspect the collected monitoring metrics and give \passfail labels on each injected failure.
Since the two benchmarks are open source and easy to follow, experienced Ph.D. students could produce accurate labels.
In case of disagreement, which turns out to be rare, they will invite industrial engineers to judge and verify difficult cases.
In particular, to address discrepancies in the impact assessment of Container CPU overload and Container memory overload failures, engineers must reconcile differences between the two Ph.D. student annotators, as the impact of these failures appears somewhat ambiguous. Resolving these discrepancies involves consulting the SLA to determine the final label. For other failures, the two PhD student annotators consistently reach an agreement.
Lastly, we convert \texttt{PASS} to $1$ and \texttt{FAIL} to $0$ before quantitatively comparing them with the resilience values.

Table~\ref{tab:dataset_statistics} shows the statistics of the three datasets.
As the number of monitoring metrics varies with the microservice system architecture, we list the number of system performance metrics (denoted as $|\mathcal{P}|$) and user-aware metrics (denoted as $|\mathcal{B}|$) in Table~\ref{tab:dataset_statistics}.
``\# Microservices'' and ``\# Failures'' mean the number of microservices, and the number of injected failures in the dataset, respectively.

\subsubsection{Baselines}

As \frmwk is the first automatic data-driven approach to compute resilience indices, few existing approaches could serve as baselines.
Since metrics are time series data and the nature of testing is classification, we resort to commonly-used classification algorithms as baselines, i.e., Support Vector Machine Classifier~\cite{DBLP:journals/ml/CortesV95} (denoted as SVC), Random Forest~\cite{DBLP:journals/ml/Breiman01} (denoted as RF), and Extra Trees~\cite{DBLP:journals/ml/GeurtsEW06} (denoted as ET).
For the baselines, we directly use the implementation from the Python package \texttt{sklearn}.

Since \frmwk does not require training, to ensure fairness, we only use the classification baselines to compute the contribution of different metrics to the overall degradation.
Specifically, let the input $X_t$ be all the monitoring data at time $t$, and the output $y_t$ be whether $t$ is in the failure injection period, we train the predictive baselines with all $X_t$ and $y_t$.
No testing data is needed, as we directly use the rank of feature importance (for ET and RF) and the rank of coefficient (for SVC) as the ordered sequence of the monitoring metrics.
In RF and ET, feature importance describes the relevance of features~\cite{DBLP:journals/ml/Breiman01}. The meaning of coefficients in SVC is in line with feature importance~\cite{platt1999probabilistic}.
As the baselines already consider the relevance of features, we set the contribution $c_i=1$ and calculate the resilience indices the same way as \frmwk.

\subsubsection{Evaluation Metrics}

Since the label is binary, but the resilience index of \frmwk is a decimal value, we employ two types of evaluation metrics.
First, we follow existing work~\cite{AID} and employ Mean Absolute Error (MAE) $MAE = \frac{\sum_{i=1}^N |y_i - p_i| }{n}$, Root Mean Squared Error (RMSE) $RMSE = \sqrt{\frac{\sum_{i=1}^N (y_i - p_i)^2}{N}}$, and Cross Entropy (CE) $CE = \frac{1}{N} \sum_{i=1}^N - [ y_i \log (p_i) + (1 - y_i) \log (1-p_i) ]$ to directly compare binary labels and decimal outputs.
Specifically, CE calculates the difference between the label and the probability distribution of the produced resilience indices.
MAE and RMSE measure the absolute and root-mean-squared differences between the produced resilience indices and the ground truth labels.
Lower CE, MAE, and RMSE values indicate better prediction results.
Second, to show the practical usage in practice, we set the resilience threshold $\tau$ to convert the decimal outputs to binary predictions, then use accuracy and f1-score as the evaluation metric.
Higher accuracy and f1-score indicate better prediction results.

\subsubsection{Experimental Environments}

We deployed the \TT benchmark in a Kubernetes cluster of two physical servers.
Both servers have 128 GB RAM and 24 CPU cores.
The \SN benchmark was deployed in a t2.2xlarge EC2 instance of AWS with 8 GB RAM and 8 CPU cores.
The \IND dataset was collected in proprietary servers in \company.
For all datasets, we run the degradation-based metric lattice search and the resilience indexing on a laptop with 4 Intel CPU cores and 8 GB RAM.

\subsection{RQ1: Effectiveness}
\label{sec:experiment:rq3}

To study the effectiveness of \frmwk, we compare its performance with the baseline models on both datasets.
For the contribution measure of \frmwk, we employ dynamic time warping (DTW)~\cite{dtw} algorithm.
Specifically, for the parameters of DTW, we set the warping window to be 5 seconds (for \TT) and 2 seconds (for \SN), and use the square of the absolute difference as the distance measure.
We do this because the \SN benchmark is deployed in a single server, and it responds faster than the \TT benchmark.
Moreover, real-world industrial data often contains noise and variability.
Setting a bigger window helps filter out irrelevant fluctuations and focus on meaningful patterns or trends within the data.
We also use the moving average of a window size 3 to smoothen the monitoring metrics for the baselines and our method.
We set the resilience threshold $\tau=0.4$ for the \TT and \SN datasets and set $\tau=0.75$ for the industrial dataset.
The thresholds differ between simulated and industrial datasets because \TT and \SN are equipped with very few resilience measures, but the industrial system has many resilience measures, such as replications and better error tolerance in code.
The overall performance is shown in Table~\ref{tab:rq3}, where we mark the best result for each metric and dataset.
\frmwk achieves the best performance on all the datasets.
Notably, compared with the best baseline, \frmwk reduces the loss by 44.3\%, 21.4\%, and 26.3\% in terms of CE, MAE, and RMSE on the industrial dataset.
Moreover, \frmwk achieves the best accuracy (0.8929) and F1-score (0.9362) on the industrial dataset.
The performance on the industrial dataset highlights the effectiveness of \frmwk in production.
The improvement of \frmwk on the industrial dataset is smaller than on the simulated datasets.
The industrial microservice system incorporates more fault tolerance mechanisms, making it harder for \frmwk to discriminate between \passfail.
Moreover, since the interactions of the TT benchmark are very fast, the statuses of TT's services are relatively similar, making simple baselines and our approach perform similarly.

\subsection{RQ2: Ablation Study}
\label{sec:experiment:rq4}

In RQ2, we focus on comparing the performance of various \frmwk variants to assess how different contribution measures affect \frmwk' overall performance.
To achieve this, we keep the primary framework, namely the Dissemination-based Metric Lattice Search, and employ diverse contribution measures for the performance evaluation.
In particular, we conduct experiments with varying contribution measures, i.e., Euclidean Distance (denoted as ``\frmwk-euc''), Pearson Correlation (denoted as ``\frmwk-corr''), Complexity Invariant Distance~\cite{cid1,cid2} (denoted as ``\frmwk-cid''), and keep other parameters identical.
Our method, which uses DTW as the contribution measure, is denoted as ``\frmwk'' in the table.
Table~\ref{tab:rq3} shows the performance under different contribution measures.
We marked models with the best performance in terms of CE, MAE, RMSE, and F1-score.
The results indicate that the impact of different contribution measures in a reasonable range is small, but ``\frmwk'' gives the overall best performance.

\subsection{RQ3: Efficiency}
\label{sec:experiment:rq5}

The efficiency of \frmwk is composed of three parts, including (1) the duration of failure execution, the time complexity of (2) the degradation-based metric lattice search, and (3) the resilience indexing.
The required duration of failure execution varies dramatically for different systems, and we will discuss the suggested duration in Section~\cref{sec:discussion}.
Among the remaining two phases, the most time-consuming phase is the degradation-based metric lattice search.
Theoretically, the time complexity of Algorithm~\ref{algo:demeter_degradation} depends on the length $T$ of the monitoring metrics, the number of monitoring metrics $|\mathcal{M}|$, and the time complexity of \texttt{Cont()}.
Since $|\mathcal{M}| << T$ in practice, we treat $|\mathcal{M}|$ as a constant.
The computation of performance difference costs $O(T)$, and the dimension reduction with PCA costs $O(T^3)$.
As dynamic time warping can be easily parallelized, we treat the time complexity of \texttt{Contribution()} as $O(T)$.
Merging together, the upper bound of the time complexity is $O(T^3)$.
Considering the average time of failure injection is usually several hours, the time complexity of $O(T^3)$ will not be a problem.
On average, the latter two phases take $302$ seconds to process a failure test case of $T=1200$ on a laptop.

\section{Successful Cases}
\label{sec:case}

\begin{figure}[ht]
    \centering
    \subfigure[Case 1: Process Killed \textbf{\texttt{(FAIL)}}]{
        \includegraphics[width=0.45\columnwidth]{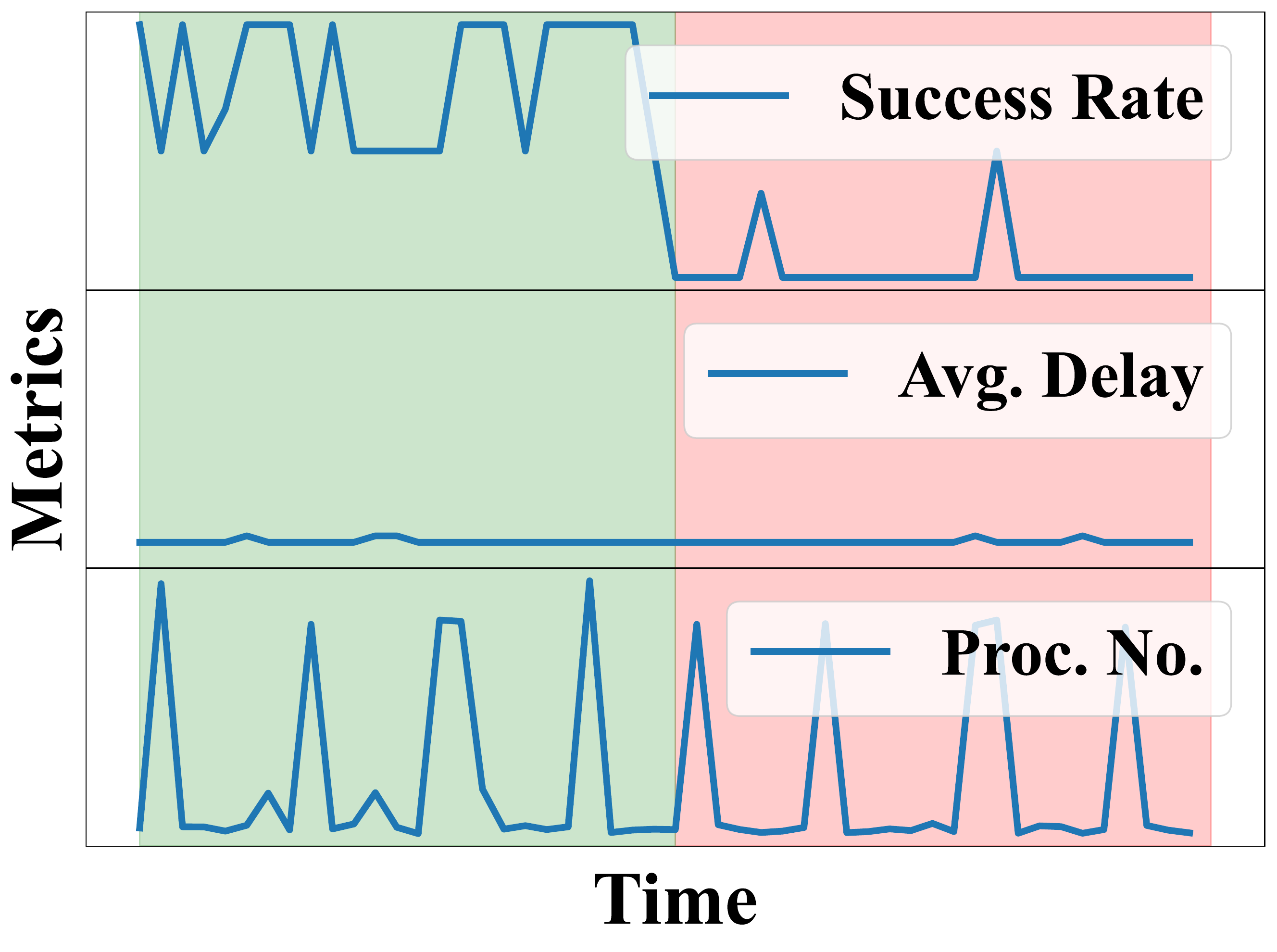}%
        \label{fig:case-kill}
    }
    \subfigure[Case 2: High I/O Rate \textbf{\texttt{({PASS})}}]{
        \includegraphics[width=0.45\columnwidth]{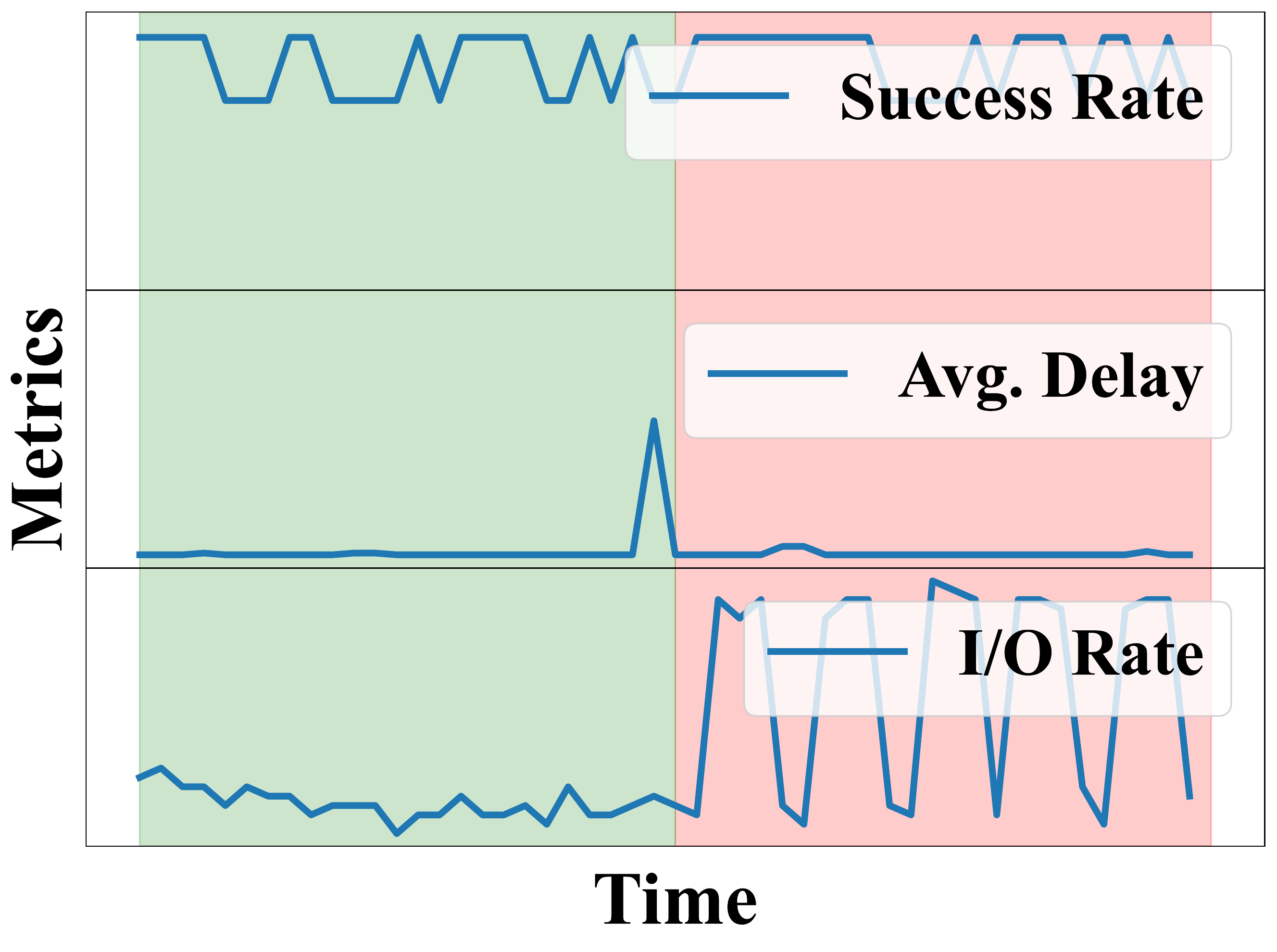}%
        \label{fig:case-io}
    }
    \vspace{-0.1in}
    \caption{Two successful cases in the industrial dataset. The \textcolor{mygreen}{green area} means the normal period and the \textcolor{myred}{red area} means the failure injection period.}
    \vspace{-0.15in}
    \label{fig:case}
\end{figure}

\frmwk has been integrated into the resilience testing procedure of \company.
This section demonstrates two cases to show the flexibility of \frmwk and the practical usage of the resilience index.
In Figure~\ref{fig:case}, \textbf{Case 1} shows the monitoring metrics during ``Process Killed'' failure in the \IND dataset.
\textbf{Case 2} shows the monitoring metrics during ``High I/O Throughput'' failure in the \IND dataset.
``Success Rate'' and ``Avg. Delay'' are user-aware metrics.
``Proc. No.'' and ``I/O Rate'' are system performance metrics.
Other unaffected system performance metrics are omitted for clear presentation.
In the first case, the killed process cannot respond to user requests.
Consequently, the success rate drops significantly, and the resilience index is low, making it {\texttt{FAIL}} \frmwk's test.
In contrast, in the second case, although the I/O rate is high, the average delay and the success rate remain stable.
Hence, the ``I/O Rate'' ranked much higher than the two user-aware metrics, and the resilience index is high, so the test result is {\texttt{PASS}}.
Note that no separate configuration is needed for the two test cases.
Thus, \frmwk can be adopted flexibly for different failures.
Moreover, by inspecting the resilience index and the ranked metrics, engineers can quickly identify the impacted metrics, showing the practicality of \frmwk.

\section{Discussion}
\label{sec:discussion}

\subsection{Practiccal Usefulness}
\label{sec:discussion:execution}

\frmwk's practicality lies in its minimal human configuration requirements.
There is no need for engineers to configure individual systems and their faults.
Instead, engineers only have to choose the desired faults and examine the generated resilience indices.
Resilience indices play a crucial role in helping engineers comprehend the extent of degradation propagation from system performance metrics to business metrics during a specific fault.
For instance, if the degradation in the network received bytes (rx\_bytes) has a more significant impact on request throughput, engineers can improve fault tolerance in the network accordingly.

Notably, \frmwk has already been integrated into the production system of \company, contributing to a reduction in the average time for resilience testing from 2 days to 4 hours.
According to our practical usage in \company, we suggest during the failure execution phase, at least 40 requests (20 requests during the failure injection and 20 requests during the normal period) should be processed.
In industrial cloud systems, a complex API request usually takes 10 seconds to finish, so the failure execution phase should last for less than 8 minutes, as a rule of thumb.
Compared with the frequency of microservice updates, i.e., usually, once a week, the failure execution phase will not burden test engineers too much with the help of \frmwk.

\subsection{Limitation}
\label{sec:discussion:limitation}
The major limitation lies in the fact that \frmwk still requires bootstrapping manual configurations on the categorization of user-aware metrics and system performance metrics to facilitate resilience indexing.
It is important to highlight, however, that these two manual tasks only need to be done once, as subsequent executions of \frmwk do not rely on human intervention.
Hence, the bootstrapping manual configurations will not hinder the practical usage of \frmwk.

\subsection{Threat to Validity}
\label{sec:discussion:threats}

\subsubsection*{Labeling Accuracy}
The major threat to validity is labeling accuracy.
To evaluate \frmwk, we conduct experiments on two simulated datasets.
The evaluation on two simulated datasets requires labeling the resilience test results, but the labels may not be 100\% accurate.
However, the resilience mechanisms and deployment environment of the benchmark systems are clear to all the annotators, so the resilience test results are straightforward.
Moreover, when disagreements arise, the annotators will consult experienced test engineers who are in charge of the resilience assurance of the cloud services of \company.
The number of inaccurate labels should be small.

\vspace{-0.05in}
\subsubsection*{Insufficiency of Simulation}
For the evaluation, we deploy open-source benchmark microservice systems to simulate real-world services.
Compared with real-world services, the open-source benchmarks do not fully consider fault tolerance, resulting in poor resilience in the simulation.
Hence, the simulated dataset may not exhibit some common attributes of real online services.
However, we deploy the benchmark microservice systems with two widely-used microservice orchestrators to show the practical usefulness of \frmwk in different environments.
We also simulate concurrent and varying user requests to mimic the real-world scenario.
Most importantly, we also employ the experiments on an industrial dataset from \company, which contains more metrics and complex degradation dissemination.
The experiment results make \frmwk stand out among the baselines.
In summary, the simulated and industrial datasets can accurately reflect the theoretical and practical superiority of \frmwk.

\section{Related Work}
\label{sec:relatedwork}

\subsubsection*{Resilience Testing of Online Services}
\label{sec:relatedwork:resilience}

To ensure the ability of the system to minimize the impact of potential failures, considerable attention has been paid to resilience testing of microservices, including model-based resilience representation and analysis~\cite{DBLP:conf/icsa/MendoncaACG20,DBLP:journals/ijseke/YinD21}, non-intrusive and automated fault injection~\cite{Gremlin, Intel-liFT, LDFI, DBLP:journals/software/BasiriBRHKRR16}, scalability resilience testing~\cite{DBLP:journals/jcloudc/AhmadA22}.
\cite{DBLP:conf/icsa/MendoncaACG20} used the PRISM probabilistic model checker to analyze the behavior of the Retry and Circuit Breaker resiliency patterns.
\cite{DBLP:journals/ijseke/YinD21} proposed a Microservice Resilience Measurement Model (MRMM) to represent the resilience requirements of MSA Systems.
\cite{LDFI} proposed a lineage-driven fault injection approach to infer whether injected faults can prevent correct outcomes by exploring historical data lineage and satisfiability testing.
\cite{Gremlin} presented a non-intrusive resilience testing framework that injects faults by manipulating the network packets between microservices.
\cite{DBLP:journals/jcloudc/AhmadA22} simulated delay latency injection to assess the fault scenario's impact on the cloud software service's scalability resilience.
\cite{LDFI, Gremlin, DBLP:journals/jcloudc/AhmadA22} all require test engineers to write test descriptions and manually check assertions.
Netflix proposed chaos engineering~\cite{DBLP:journals/software/BasiriBRHKRR16} to inject faults in the system randomly.
A recent study~\cite{Intel-liFT} proposes automatically generating resilience test cases by inferring whether the injected faults can result in severe failures.

Instead of resilience profiling, most prior works mainly concern fault injection into microservices with minimal system intrusion.
Additionally, the existing approaches rely heavily on human labor or historical cases, making them less practical in cloud-scale service systems with high dynamism and complex failure models.
In contrast, \frmwk primarily emphasizes resilience profiling with minimal manual configuration, which is dispensed with historical testing cases.

\subsubsection*{Combination Searching}
\label{sec:relatedwork:searching}

Many combination searching techniques~\cite{Kendall, DBLP:journals/cssc/BargagliottiG15, Fisher} have shown its promise in reducing information redundancy and enhancing the performances of data-driven models.
The combination searching approaches fall into three categories: score-based~\cite{Kendall} and embedding-based~\cite{LASSO, SVMSelect, RandomForest, ETree}, and wrapper-based~\cite{DBLP:journals/kbs/BermejoOGP12, DBLP:conf/gecco/MostertME18, DBLP:conf/gecco/NeshatianZ09, DBLP:conf/iotdcc/RazaQ16}.
Specifically, wrapper-based approaches use different combinations of features to train the same downstream model.
Some heuristic approaches~\cite{DBLP:journals/kbs/BermejoOGP12, DBLP:conf/iotdcc/RazaQ16} have also been developed to narrow the searching space due to the high searching complexity.
\frmwk employs a wrapper-based method and overcomes the demerits of high computation cost and over-fitting by proper pruning.

\section{Conclusion}\label{sec:conclusion}
This paper intends to mitigate the labor-intensity and flexibility issues of the current practice of resilience profiling that relies on manually making rules.
We propose the first versatile resilience profiling framework, \frmwk, for microservice systems via degradation dissemination indexing.
Our insight behind \frmwk, motivated by the investigation on the impact of common failures, is that resilient deployment can effectively prevent the dissemination of degradation from system performance metrics to user-aware metrics.
\frmwk quantifies the dissemination of degradation from system performance metrics to user-aware metrics, which is a one-size-fits-all solution without architecture knowledge, thereby adaptable to different systems.
Evaluations on open-source and industrial microservice systems show that \frmwk can accurately and efficiently measure the resilience of microservice systems, outperforming all baseline methods in terms of cross-entropy.

\begin{acks}
  The work was supported by the Guangdong Key Research Program (No. 2020B010165002), the Research Grants Council of the Hong Kong Special Administrative Region, China (CUHK 14206921), and the National Natural Science Foundation of China (No. 62202511).
\end{acks}

\normalem
\bibliographystyle{ACM-Reference-Format}
\bibliography{references}


\begin{thebibliography}{58}


\ifx \showCODEN    \undefined \def \showCODEN     #1{\unskip}     \fi
\ifx \showDOI      \undefined \def \showDOI       #1{#1}\fi
\ifx \showISBNx    \undefined \def \showISBNx     #1{\unskip}     \fi
\ifx \showISBNxiii \undefined \def \showISBNxiii  #1{\unskip}     \fi
\ifx \showISSN     \undefined \def \showISSN      #1{\unskip}     \fi
\ifx \showLCCN     \undefined \def \showLCCN      #1{\unskip}     \fi
\ifx \shownote     \undefined \def \shownote      #1{#1}          \fi
\ifx \showarticletitle \undefined \def \showarticletitle #1{#1}   \fi
\ifx \showURL      \undefined \def \showURL       {\relax}        \fi
\providecommand\bibfield[2]{#2}
\providecommand\bibinfo[2]{#2}
\providecommand\natexlab[1]{#1}
\providecommand\showeprint[2][]{arXiv:#2}

\bibitem[Ahmad and Andras(2022)]%
        {DBLP:journals/jcloudc/AhmadA22}
\bibfield{author}{\bibinfo{person}{Amro~Al{-}Said Ahmad} {and}
  \bibinfo{person}{Peter Andras}.} \bibinfo{year}{2022}\natexlab{}.
\newblock \showarticletitle{Scalability resilience framework using
  application-level fault injection for cloud-based software services}.
\newblock \bibinfo{journal}{\emph{J. Cloud Comput.}}  \bibinfo{volume}{11}
  (\bibinfo{year}{2022}), \bibinfo{pages}{1}.
\newblock
\urldef\tempurl%
\url{https://doi.org/10.1186/S13677-021-00277-Z}
\showDOI{\tempurl}


\bibitem[Alibaba(2022)]%
        {chaosblade}
\bibfield{author}{\bibinfo{person}{Alibaba}.} \bibinfo{year}{2022}\natexlab{}.
\newblock \bibinfo{title}{ChaosBlade: An easy to use and powerful chaos
  engineering experiment toolkit}.
\newblock
\newblock
\urldef\tempurl%
\url{https://github.com/chaosblade-io/chaosblade}
\showURL{%
\tempurl}


\bibitem[Alvaro et~al\mbox{.}(2015)]%
        {LDFI}
\bibfield{author}{\bibinfo{person}{Peter Alvaro}, \bibinfo{person}{Joshua
  Rosen}, {and} \bibinfo{person}{Joseph~M. Hellerstein}.}
  \bibinfo{year}{2015}\natexlab{}.
\newblock \showarticletitle{Lineage-driven Fault Injection}. In
  \bibinfo{booktitle}{\emph{Proceedings of the 2015 {ACM} {SIGMOD}
  International Conference on Management of Data, Melbourne, Victoria,
  Australia, May 31 - June 4, 2015}}. \bibinfo{publisher}{{ACM}},
  \bibinfo{pages}{331--346}.
\newblock
\urldef\tempurl%
\url{https://doi.org/10.1145/2723372.2723711}
\showDOI{\tempurl}


\bibitem[Ammann and Offutt(2008)]%
        {ammann2016introduction}
\bibfield{author}{\bibinfo{person}{Paul Ammann} {and} \bibinfo{person}{Jeff
  Offutt}.} \bibinfo{year}{2008}\natexlab{}.
\newblock \bibinfo{booktitle}{\emph{Introduction to Software Testing}}.
\newblock \bibinfo{publisher}{Cambridge University Press}.
\newblock
\showISBNx{978-0-521-88038-1}
\urldef\tempurl%
\url{https://doi.org/10.1017/CBO9780511809163}
\showDOI{\tempurl}


\bibitem[Armbrust et~al\mbox{.}(2009)]%
        {berkeley-view-cloud}
\bibfield{author}{\bibinfo{person}{Michael Armbrust}, \bibinfo{person}{Armando
  Fox}, \bibinfo{person}{Rean Griffith}, \bibinfo{person}{Anthony~D. Joseph},
  \bibinfo{person}{Randy~H. Katz}, \bibinfo{person}{Andrew Konwinski},
  \bibinfo{person}{Gunho Lee}, \bibinfo{person}{David~A. Patterson},
  \bibinfo{person}{Ariel Rabkin}, \bibinfo{person}{Ion Stoica}, {and}
  \bibinfo{person}{Matei Zaharia}.} \bibinfo{year}{2009}\natexlab{}.
\newblock \bibinfo{booktitle}{\emph{Above the Clouds: A Berkeley View of Cloud
  Computing}}.
\newblock \bibinfo{type}{{T}echnical {R}eport} UCB/EECS-2009-28.
  \bibinfo{institution}{EECS Department, University of California, Berkeley}.
\newblock
\urldef\tempurl%
\url{http://www2.eecs.berkeley.edu/Pubs/TechRpts/2009/EECS-2009-28.html}
\showURL{%
\tempurl}


\bibitem[Bargagliotti and Greenwell(2015)]%
        {DBLP:journals/cssc/BargagliottiG15}
\bibfield{author}{\bibinfo{person}{Anna~E. Bargagliotti} {and}
  \bibinfo{person}{Raymond~N. Greenwell}.} \bibinfo{year}{2015}\natexlab{}.
\newblock \showarticletitle{Combinatorics and Statistical Issues Related to the
  Kruskal-Wallis Statistic}.
\newblock \bibinfo{journal}{\emph{Commun. Stat. Simul. Comput.}}
  \bibinfo{volume}{44}, \bibinfo{number}{2} (\bibinfo{year}{2015}),
  \bibinfo{pages}{533--550}.
\newblock
\urldef\tempurl%
\url{https://doi.org/10.1080/03610918.2013.786781}
\showURL{%
\tempurl}


\bibitem[Basiri et~al\mbox{.}(2016a)]%
        {basiri2016chaos}
\bibfield{author}{\bibinfo{person}{Ali Basiri}, \bibinfo{person}{Niosha
  Behnam}, \bibinfo{person}{Ruud de Rooij}, \bibinfo{person}{Lorin Hochstein},
  \bibinfo{person}{Luke Kosewski}, \bibinfo{person}{Justin Reynolds}, {and}
  \bibinfo{person}{Casey Rosenthal}.} \bibinfo{year}{2016}\natexlab{a}.
\newblock \showarticletitle{Chaos Engineering}.
\newblock \bibinfo{journal}{\emph{{IEEE} Softw.}} \bibinfo{volume}{33},
  \bibinfo{number}{3} (\bibinfo{year}{2016}), \bibinfo{pages}{35--41}.
\newblock
\urldef\tempurl%
\url{https://doi.org/10.1109/MS.2016.60}
\showDOI{\tempurl}


\bibitem[Basiri et~al\mbox{.}(2016b)]%
        {DBLP:journals/software/BasiriBRHKRR16}
\bibfield{author}{\bibinfo{person}{Ali Basiri}, \bibinfo{person}{Niosha
  Behnam}, \bibinfo{person}{Ruud de Rooij}, \bibinfo{person}{Lorin Hochstein},
  \bibinfo{person}{Luke Kosewski}, \bibinfo{person}{Justin Reynolds}, {and}
  \bibinfo{person}{Casey Rosenthal}.} \bibinfo{year}{2016}\natexlab{b}.
\newblock \showarticletitle{Chaos Engineering}.
\newblock \bibinfo{journal}{\emph{{IEEE} Softw.}} \bibinfo{volume}{33},
  \bibinfo{number}{3} (\bibinfo{year}{2016}), \bibinfo{pages}{35--41}.
\newblock
\urldef\tempurl%
\url{https://doi.org/10.1109/MS.2016.60}
\showDOI{\tempurl}


\bibitem[Batista et~al\mbox{.}(2014)]%
        {cid1}
\bibfield{author}{\bibinfo{person}{Gustavo E. A. P.~A. Batista},
  \bibinfo{person}{Eamonn~J. Keogh}, \bibinfo{person}{Oben~Moses Tataw}, {and}
  \bibinfo{person}{Vin{\'{\i}}cius M.~A. de Souza}.}
  \bibinfo{year}{2014}\natexlab{}.
\newblock \showarticletitle{{CID:} an efficient complexity-invariant distance
  for time series}.
\newblock \bibinfo{journal}{\emph{Data Min. Knowl. Discov.}}
  \bibinfo{volume}{28}, \bibinfo{number}{3} (\bibinfo{year}{2014}),
  \bibinfo{pages}{634--669}.
\newblock
\urldef\tempurl%
\url{https://doi.org/10.1007/S10618-013-0312-3}
\showDOI{\tempurl}


\bibitem[Batista et~al\mbox{.}(2011)]%
        {cid2}
\bibfield{author}{\bibinfo{person}{Gustavo E. A. P.~A. Batista},
  \bibinfo{person}{Xiaoyue Wang}, {and} \bibinfo{person}{Eamonn~J. Keogh}.}
  \bibinfo{year}{2011}\natexlab{}.
\newblock \showarticletitle{A Complexity-Invariant Distance Measure for Time
  Series}. In \bibinfo{booktitle}{\emph{Proceedings of the Eleventh {SIAM}
  International Conference on Data Mining, {SDM} 2011, April 28-30, 2011, Mesa,
  Arizona, {USA}}}. \bibinfo{publisher}{{SIAM} / Omnipress},
  \bibinfo{pages}{699--710}.
\newblock
\urldef\tempurl%
\url{https://doi.org/10.1137/1.9781611972818.60}
\showDOI{\tempurl}


\bibitem[Bermejo et~al\mbox{.}(2012)]%
        {DBLP:journals/kbs/BermejoOGP12}
\bibfield{author}{\bibinfo{person}{Pablo Bermejo}, \bibinfo{person}{Luis de~la
  Ossa}, \bibinfo{person}{Jos{\'{e}}~A. G{\'{a}}mez}, {and}
  \bibinfo{person}{Jos{\'{e}}~Miguel Puerta}.} \bibinfo{year}{2012}\natexlab{}.
\newblock \showarticletitle{Fast wrapper feature subset selection in
  high-dimensional datasets by means of filter re-ranking}.
\newblock \bibinfo{journal}{\emph{Knowl. Based Syst.}} \bibinfo{volume}{25},
  \bibinfo{number}{1} (\bibinfo{year}{2012}), \bibinfo{pages}{35--44}.
\newblock
\urldef\tempurl%
\url{https://doi.org/10.1016/J.KNOSYS.2011.01.015}
\showDOI{\tempurl}


\bibitem[Blohowiak et~al\mbox{.}(2016)]%
        {DBLP:conf/issre/BlohowiakBHR16}
\bibfield{author}{\bibinfo{person}{Aaron Blohowiak}, \bibinfo{person}{Ali
  Basiri}, \bibinfo{person}{Lorin Hochstein}, {and} \bibinfo{person}{Casey
  Rosenthal}.} \bibinfo{year}{2016}\natexlab{}.
\newblock \showarticletitle{A Platform for Automating Chaos Experiments}. In
  \bibinfo{booktitle}{\emph{2016 {IEEE} International Symposium on Software
  Reliability Engineering Workshops, {ISSRE} Workshops 2016, Ottawa, ON,
  Canada, October 23-27, 2016}}. \bibinfo{publisher}{{IEEE} Computer Society},
  \bibinfo{pages}{5--8}.
\newblock
\urldef\tempurl%
\url{https://doi.org/10.1109/ISSREW.2016.52}
\showDOI{\tempurl}


\bibitem[Breiman(2001)]%
        {DBLP:journals/ml/Breiman01}
\bibfield{author}{\bibinfo{person}{Leo Breiman}.}
  \bibinfo{year}{2001}\natexlab{}.
\newblock \showarticletitle{Random Forests}.
\newblock \bibinfo{journal}{\emph{Mach. Learn.}} \bibinfo{volume}{45},
  \bibinfo{number}{1} (\bibinfo{year}{2001}), \bibinfo{pages}{5--32}.
\newblock
\urldef\tempurl%
\url{https://doi.org/10.1023/A:1010933404324}
\showDOI{\tempurl}


\bibitem[Chadwick and Kurz(1969)]%
        {Kendall}
\bibfield{author}{\bibinfo{person}{Henry~D. Chadwick} {and}
  \bibinfo{person}{Ludwik Kurz}.} \bibinfo{year}{1969}\natexlab{}.
\newblock \showarticletitle{Rank permutation group codes based on Kendall's
  correlation statistic}.
\newblock \bibinfo{journal}{\emph{{IEEE} Trans. Inf. Theory}}
  \bibinfo{volume}{15}, \bibinfo{number}{2} (\bibinfo{year}{1969}),
  \bibinfo{pages}{306--315}.
\newblock
\urldef\tempurl%
\url{https://doi.org/10.1109/TIT.1969.1054291}
\showDOI{\tempurl}


\bibitem[Cortes and Vapnik(1995)]%
        {DBLP:journals/ml/CortesV95}
\bibfield{author}{\bibinfo{person}{Corinna Cortes} {and}
  \bibinfo{person}{Vladimir Vapnik}.} \bibinfo{year}{1995}\natexlab{}.
\newblock \showarticletitle{Support-Vector Networks}.
\newblock \bibinfo{journal}{\emph{Mach. Learn.}} \bibinfo{volume}{20},
  \bibinfo{number}{3} (\bibinfo{year}{1995}), \bibinfo{pages}{273--297}.
\newblock
\urldef\tempurl%
\url{https://doi.org/10.1007/BF00994018}
\showDOI{\tempurl}


\bibitem[Croft et~al\mbox{.}(2010)]%
        {croft2010search}
\bibfield{author}{\bibinfo{person}{W~Bruce Croft}, \bibinfo{person}{Donald
  Metzler}, {and} \bibinfo{person}{Trevor Strohman}.}
  \bibinfo{year}{2010}\natexlab{}.
\newblock \bibinfo{booktitle}{\emph{Search engines: Information retrieval in
  practice}}. Vol.~\bibinfo{volume}{520}.
\newblock \bibinfo{publisher}{Addison-Wesley Reading}.
\newblock


\bibitem[Dickson(2013)]%
        {theory-of-monitoring}
\bibfield{author}{\bibinfo{person}{Caskey~L. Dickson}.}
  \bibinfo{year}{2013}\natexlab{}.
\newblock \bibinfo{booktitle}{\emph{A Working Theory-of-Monitoring}}.
\newblock \bibinfo{type}{{T}echnical {R}eport}. \bibinfo{institution}{Google,
  Inc.}
\newblock
\urldef\tempurl%
\url{https://www.usenix.org/conference/lisa13/working-theory-monitoring}
\showURL{%
\tempurl}


\bibitem[Dillon and Haimes(1996)]%
        {ETree}
\bibfield{author}{\bibinfo{person}{Robin Dillon} {and}
  \bibinfo{person}{Yacov~Y. Haimes}.} \bibinfo{year}{1996}\natexlab{}.
\newblock \showarticletitle{Risk of extreme events via multiobjective decision
  trees: application to telecommunications}.
\newblock \bibinfo{journal}{\emph{{IEEE} Trans. Syst. Man Cybern. Part {A}}}
  \bibinfo{volume}{26}, \bibinfo{number}{2} (\bibinfo{year}{1996}),
  \bibinfo{pages}{262--271}.
\newblock
\urldef\tempurl%
\url{https://doi.org/10.1109/3468.485753}
\showURL{%
\tempurl}


\bibitem[Doc(2019)]%
        {microservice-architecture}
\bibfield{author}{\bibinfo{person}{Microsoft~Azure Doc}.}
  \bibinfo{year}{2019}\natexlab{}.
\newblock \bibinfo{title}{Microservices architecture style}.
\newblock
\newblock
\urldef\tempurl%
\url{https://docs.microsoft.com/en-us/azure/architecture/guide/architecture-styles/microservices}
\showURL{%
\tempurl}


\bibitem[Foundation(2022a)]%
        {kubernetes}
\bibfield{author}{\bibinfo{person}{The~Linux Foundation}.}
  \bibinfo{year}{2022}\natexlab{a}.
\newblock \bibinfo{title}{Kubernetes}.
\newblock
\newblock
\urldef\tempurl%
\url{http://kubernetes.io/}
\showURL{%
\tempurl}


\bibitem[Foundation(2022b)]%
        {prometheus}
\bibfield{author}{\bibinfo{person}{The~Linux Foundation}.}
  \bibinfo{year}{2022}\natexlab{b}.
\newblock \bibinfo{title}{Prometheus}.
\newblock
\newblock
\urldef\tempurl%
\url{https://prometheus.io/}
\showURL{%
\tempurl}


\bibitem[Gabrielson(2022)]%
        {aws-challenges-with-distributed-systems}
\bibfield{author}{\bibinfo{person}{Jacob Gabrielson}.}
  \bibinfo{year}{2022}\natexlab{}.
\newblock \bibinfo{title}{Challenges with distributed systems}.
\newblock
\newblock
\urldef\tempurl%
\url{https://aws.amazon.com/builders-library/challenges-with-distributed-systems/}
\showURL{%
\tempurl}


\bibitem[Gan et~al\mbox{.}(2021)]%
        {DBLP:conf/asplos/0002LD0D21}
\bibfield{author}{\bibinfo{person}{Yu Gan}, \bibinfo{person}{Mingyu Liang},
  \bibinfo{person}{Sundar Dev}, \bibinfo{person}{David Lo}, {and}
  \bibinfo{person}{Christina Delimitrou}.} \bibinfo{year}{2021}\natexlab{}.
\newblock \showarticletitle{Sage: practical and scalable ML-driven performance
  debugging in microservices}. In \bibinfo{booktitle}{\emph{{ASPLOS} '21: 26th
  {ACM} International Conference on Architectural Support for Programming
  Languages and Operating Systems, Virtual Event, USA, April 19-23, 2021}}.
  \bibinfo{publisher}{{ACM}}, \bibinfo{pages}{135--151}.
\newblock
\urldef\tempurl%
\url{https://doi.org/10.1145/3445814.3446700}
\showURL{%
\tempurl}


\bibitem[Gan et~al\mbox{.}(2019a)]%
        {DBLP:conf/asplos/GanZCSRKBHRJHPH19}
\bibfield{author}{\bibinfo{person}{Yu Gan}, \bibinfo{person}{Yanqi Zhang},
  \bibinfo{person}{Dailun Cheng}, \bibinfo{person}{Ankitha Shetty},
  \bibinfo{person}{Priyal Rathi}, \bibinfo{person}{Nayan Katarki},
  \bibinfo{person}{Ariana Bruno}, \bibinfo{person}{Justin Hu},
  \bibinfo{person}{Brian Ritchken}, \bibinfo{person}{Brendon Jackson},
  \bibinfo{person}{Kelvin Hu}, \bibinfo{person}{Meghna Pancholi},
  \bibinfo{person}{Yuan He}, \bibinfo{person}{Brett Clancy},
  \bibinfo{person}{Chris Colen}, \bibinfo{person}{Fukang Wen},
  \bibinfo{person}{Catherine Leung}, \bibinfo{person}{Siyuan Wang},
  \bibinfo{person}{Leon Zaruvinsky}, \bibinfo{person}{Mateo Espinosa},
  \bibinfo{person}{Rick Lin}, \bibinfo{person}{Zhongling Liu},
  \bibinfo{person}{Jake Padilla}, {and} \bibinfo{person}{Christina
  Delimitrou}.} \bibinfo{year}{2019}\natexlab{a}.
\newblock \showarticletitle{An Open-Source Benchmark Suite for Microservices
  and Their Hardware-Software Implications for Cloud {\&} Edge Systems}. In
  \bibinfo{booktitle}{\emph{Proceedings of the Twenty-Fourth International
  Conference on Architectural Support for Programming Languages and Operating
  Systems, {ASPLOS} 2019, Providence, RI, USA, April 13-17, 2019}}.
  \bibinfo{publisher}{{ACM}}, \bibinfo{pages}{3--18}.
\newblock
\urldef\tempurl%
\url{https://doi.org/10.1145/3297858.3304013}
\showDOI{\tempurl}


\bibitem[Gan et~al\mbox{.}(2019b)]%
        {DBLP:conf/asplos/GanZHCHPD19}
\bibfield{author}{\bibinfo{person}{Yu Gan}, \bibinfo{person}{Yanqi Zhang},
  \bibinfo{person}{Kelvin Hu}, \bibinfo{person}{Dailun Cheng},
  \bibinfo{person}{Yuan He}, \bibinfo{person}{Meghna Pancholi}, {and}
  \bibinfo{person}{Christina Delimitrou}.} \bibinfo{year}{2019}\natexlab{b}.
\newblock \showarticletitle{Seer: Leveraging Big Data to Navigate the
  Complexity of Performance Debugging in Cloud Microservices}. In
  \bibinfo{booktitle}{\emph{Proceedings of the Twenty-Fourth International
  Conference on Architectural Support for Programming Languages and Operating
  Systems, {ASPLOS} 2019, Providence, RI, USA, April 13-17, 2019}}.
  \bibinfo{publisher}{{ACM}}, \bibinfo{pages}{19--33}.
\newblock
\urldef\tempurl%
\url{https://doi.org/10.1145/3297858.3304004}
\showDOI{\tempurl}


\bibitem[Geurts et~al\mbox{.}(2006)]%
        {DBLP:journals/ml/GeurtsEW06}
\bibfield{author}{\bibinfo{person}{Pierre Geurts}, \bibinfo{person}{Damien
  Ernst}, {and} \bibinfo{person}{Louis Wehenkel}.}
  \bibinfo{year}{2006}\natexlab{}.
\newblock \showarticletitle{Extremely randomized trees}.
\newblock \bibinfo{journal}{\emph{Mach. Learn.}} \bibinfo{volume}{63},
  \bibinfo{number}{1} (\bibinfo{year}{2006}), \bibinfo{pages}{3--42}.
\newblock
\urldef\tempurl%
\url{https://doi.org/10.1007/S10994-006-6226-1}
\showDOI{\tempurl}


\bibitem[Google(2022)]%
        {cadvisor}
\bibfield{author}{\bibinfo{person}{Google}.} \bibinfo{year}{2022}\natexlab{}.
\newblock \bibinfo{title}{cAdvisor: Analyzes resource usage and performance
  characteristics of running containers.}
\newblock
\newblock
\urldef\tempurl%
\url{https://github.com/google/cadvisor}
\showURL{%
\tempurl}


\bibitem[Gu et~al\mbox{.}(2011)]%
        {Fisher}
\bibfield{author}{\bibinfo{person}{Quanquan Gu}, \bibinfo{person}{Zhenhui Li},
  {and} \bibinfo{person}{Jiawei Han}.} \bibinfo{year}{2011}\natexlab{}.
\newblock \showarticletitle{Generalized Fisher Score for Feature Selection}. In
  \bibinfo{booktitle}{\emph{{UAI} 2011, Proceedings of the Twenty-Seventh
  Conference on Uncertainty in Artificial Intelligence, Barcelona, Spain, July
  14-17, 2011}}. \bibinfo{publisher}{{AUAI} Press}, \bibinfo{pages}{266--273}.
\newblock
\urldef\tempurl%
\url{https://dslpitt.org/uai/displayArticleDetails.jsp?mmnu=1\&smnu=2\&article\_id=2175\&proceeding\_id=27}
\showURL{%
\tempurl}


\bibitem[Gunawi et~al\mbox{.}(2016)]%
        {DBLP:conf/cloud/GunawiHSLSAE16}
\bibfield{author}{\bibinfo{person}{Haryadi~S. Gunawi}, \bibinfo{person}{Mingzhe
  Hao}, \bibinfo{person}{Riza~O. Suminto}, \bibinfo{person}{Agung Laksono},
  \bibinfo{person}{Anang~D. Satria}, \bibinfo{person}{Jeffry Adityatama}, {and}
  \bibinfo{person}{Kurnia~J. Eliazar}.} \bibinfo{year}{2016}\natexlab{}.
\newblock \showarticletitle{Why Does the Cloud Stop Computing? Lessons from
  Hundreds of Service Outages}. In \bibinfo{booktitle}{\emph{Proceedings of the
  Seventh {ACM} Symposium on Cloud Computing, Santa Clara, CA, USA, October
  5-7, 2016}}. \bibinfo{publisher}{{ACM}}, \bibinfo{pages}{1--16}.
\newblock
\urldef\tempurl%
\url{https://doi.org/10.1145/2987550.2987583}
\showDOI{\tempurl}


\bibitem[Han et~al\mbox{.}(2011)]%
        {han2011data}
\bibfield{author}{\bibinfo{person}{Jiawei Han}, \bibinfo{person}{Micheline
  Kamber}, {and} \bibinfo{person}{Jian Pei}.} \bibinfo{year}{2011}\natexlab{}.
\newblock \bibinfo{booktitle}{\emph{Data Mining: Concepts and Techniques, 3rd
  edition}}.
\newblock \bibinfo{publisher}{Morgan Kaufmann}.
\newblock
\showISBNx{978-0123814791}
\urldef\tempurl%
\url{http://hanj.cs.illinois.edu/bk3/}
\showURL{%
\tempurl}


\bibitem[Heorhiadi et~al\mbox{.}(2016)]%
        {Gremlin}
\bibfield{author}{\bibinfo{person}{Victor Heorhiadi}, \bibinfo{person}{Shriram
  Rajagopalan}, \bibinfo{person}{Hani Jamjoom}, \bibinfo{person}{Michael~K.
  Reiter}, {and} \bibinfo{person}{Vyas Sekar}.}
  \bibinfo{year}{2016}\natexlab{}.
\newblock \showarticletitle{Gremlin: Systematic Resilience Testing of
  Microservices}. In \bibinfo{booktitle}{\emph{36th {IEEE} International
  Conference on Distributed Computing Systems, {ICDCS} 2016, Nara, Japan, June
  27-30, 2016}}. \bibinfo{publisher}{{IEEE} Computer Society},
  \bibinfo{pages}{57--66}.
\newblock
\urldef\tempurl%
\url{https://doi.org/10.1109/ICDCS.2016.11}
\showDOI{\tempurl}


\bibitem[Ho(1998)]%
        {RandomForest}
\bibfield{author}{\bibinfo{person}{Tin~Kam Ho}.}
  \bibinfo{year}{1998}\natexlab{}.
\newblock \showarticletitle{The Random Subspace Method for Constructing
  Decision Forests}.
\newblock \bibinfo{journal}{\emph{{IEEE} Trans. Pattern Anal. Mach. Intell.}}
  \bibinfo{volume}{20}, \bibinfo{number}{8} (\bibinfo{year}{1998}),
  \bibinfo{pages}{832--844}.
\newblock
\urldef\tempurl%
\url{https://doi.org/10.1109/34.709601}
\showURL{%
\tempurl}


\bibitem[Huang et~al\mbox{.}(2017)]%
        {gray-failure}
\bibfield{author}{\bibinfo{person}{Peng Huang}, \bibinfo{person}{Chuanxiong
  Guo}, \bibinfo{person}{Lidong Zhou}, \bibinfo{person}{Jacob~R. Lorch},
  \bibinfo{person}{Yingnong Dang}, \bibinfo{person}{Murali Chintalapati}, {and}
  \bibinfo{person}{Randolph Yao}.} \bibinfo{year}{2017}\natexlab{}.
\newblock \showarticletitle{Gray Failure: The Achilles' Heel of Cloud-Scale
  Systems}. In \bibinfo{booktitle}{\emph{Proceedings of the 16th Workshop on
  Hot Topics in Operating Systems, HotOS 2017, Whistler, BC, Canada, May 8-10,
  2017}}. \bibinfo{publisher}{{ACM}}, \bibinfo{pages}{150--155}.
\newblock
\urldef\tempurl%
\url{https://doi.org/10.1145/3102980.3103005}
\showDOI{\tempurl}


\bibitem[Jagadeesan and Mendiratta(2020)]%
        {DBLP:conf/issre/JagadeesanM20}
\bibfield{author}{\bibinfo{person}{Lalita~Jategaonkar Jagadeesan} {and}
  \bibinfo{person}{Veena~B. Mendiratta}.} \bibinfo{year}{2020}\natexlab{}.
\newblock \showarticletitle{When Failure is (Not) an Option: Reliability Models
  for Microservices Architectures}. In \bibinfo{booktitle}{\emph{2020 {IEEE}
  International Symposium on Software Reliability Engineering Workshops,
  {ISSRE} Workshops, Coimbra, Portugal, October 12-15, 2020}}.
  \bibinfo{publisher}{{IEEE}}, \bibinfo{pages}{19--24}.
\newblock
\urldef\tempurl%
\url{https://doi.org/10.1109/ISSREW51248.2020.00031}
\showDOI{\tempurl}


\bibitem[Keogh(2002)]%
        {dtw}
\bibfield{author}{\bibinfo{person}{Eamonn~J. Keogh}.}
  \bibinfo{year}{2002}\natexlab{}.
\newblock \showarticletitle{Exact Indexing of Dynamic Time Warping}. In
  \bibinfo{booktitle}{\emph{Proceedings of 28th International Conference on
  Very Large Data Bases, {VLDB} 2002, Hong Kong, August 20-23, 2002}}.
  \bibinfo{publisher}{Morgan Kaufmann}, \bibinfo{pages}{406--417}.
\newblock
\urldef\tempurl%
\url{https://doi.org/10.1016/B978-155860869-6/50043-3}
\showDOI{\tempurl}


\bibitem[Kim and Kim(2004)]%
        {LASSO}
\bibfield{author}{\bibinfo{person}{Yongdai Kim} {and} \bibinfo{person}{Jinseog
  Kim}.} \bibinfo{year}{2004}\natexlab{}.
\newblock \showarticletitle{Gradient {LASSO} for feature selection}. In
  \bibinfo{booktitle}{\emph{Machine Learning, Proceedings of the Twenty-first
  International Conference {(ICML} 2004), Banff, Alberta, Canada, July 4-8,
  2004}} \emph{(\bibinfo{series}{{ACM} International Conference Proceeding
  Series}, Vol.~\bibinfo{volume}{69})}. \bibinfo{publisher}{{ACM}}.
\newblock
\urldef\tempurl%
\url{https://doi.org/10.1145/1015330.1015364}
\showURL{%
\tempurl}


\bibitem[Kubernetes(2022a)]%
        {k8s-doc-arch}
\bibfield{author}{\bibinfo{person}{Kubernetes}.}
  \bibinfo{year}{2022}\natexlab{a}.
\newblock \bibinfo{title}{Kubernetes Documentation: Cluster Architecture}.
\newblock
\newblock
\urldef\tempurl%
\url{https://kubernetes.io/docs/concepts/architecture/}
\showURL{%
\tempurl}


\bibitem[Kubernetes(2022b)]%
        {k8s-doc-disruption}
\bibfield{author}{\bibinfo{person}{Kubernetes}.}
  \bibinfo{year}{2022}\natexlab{b}.
\newblock \bibinfo{title}{Kubernetes Documentation: Disruptions}.
\newblock
\newblock
\urldef\tempurl%
\url{https://kubernetes.io/docs/concepts/workloads/pods/disruptions/}
\showURL{%
\tempurl}


\bibitem[Lewis(2021)]%
        {techtarget-blog}
\bibfield{author}{\bibinfo{person}{Sarah Lewis}.}
  \bibinfo{year}{2021}\natexlab{}.
\newblock \bibinfo{title}{Software Resilience Testing}.
\newblock
\newblock
\urldef\tempurl%
\url{https://www.techtarget.com/searchsoftwarequality/definition/software-resilience-testing}
\showURL{%
\tempurl}


\bibitem[Liu et~al\mbox{.}(2019)]%
        {DBLP:conf/hotos/LiuLMN19}
\bibfield{author}{\bibinfo{person}{Haopeng Liu}, \bibinfo{person}{Shan Lu},
  \bibinfo{person}{Madan Musuvathi}, {and} \bibinfo{person}{Suman Nath}.}
  \bibinfo{year}{2019}\natexlab{}.
\newblock \showarticletitle{What bugs cause production cloud incidents?}. In
  \bibinfo{booktitle}{\emph{Proceedings of the Workshop on Hot Topics in
  Operating Systems, HotOS 2019, Bertinoro, Italy, May 13-15, 2019}}.
  \bibinfo{publisher}{{ACM}}, \bibinfo{pages}{155--162}.
\newblock
\urldef\tempurl%
\url{https://doi.org/10.1145/3317550.3321438}
\showDOI{\tempurl}


\bibitem[Long et~al\mbox{.}(2020)]%
        {Intel-liFT}
\bibfield{author}{\bibinfo{person}{Zhenyue Long}, \bibinfo{person}{Guoquan Wu},
  \bibinfo{person}{Xiaojiang Chen}, \bibinfo{person}{Chengxu Cui},
  \bibinfo{person}{Wei Chen}, {and} \bibinfo{person}{Jun Wei}.}
  \bibinfo{year}{2020}\natexlab{}.
\newblock \showarticletitle{Fitness-guided Resilience Testing of
  Microservice-based Applications}. In \bibinfo{booktitle}{\emph{2020 {IEEE}
  International Conference on Web Services, {ICWS} 2020, Beijing, China,
  October 19-23, 2020}}. \bibinfo{publisher}{{IEEE}},
  \bibinfo{pages}{151--158}.
\newblock
\urldef\tempurl%
\url{https://doi.org/10.1109/ICWS49710.2020.00027}
\showDOI{\tempurl}


\bibitem[Maldonado and L{\'{o}}pez(2018)]%
        {SVMSelect}
\bibfield{author}{\bibinfo{person}{Sebasti{\'{a}}n Maldonado} {and}
  \bibinfo{person}{Julio L{\'{o}}pez}.} \bibinfo{year}{2018}\natexlab{}.
\newblock \showarticletitle{Dealing with high-dimensional class-imbalanced
  datasets: Embedded feature selection for {SVM} classification}.
\newblock \bibinfo{journal}{\emph{Appl. Soft Comput.}}  \bibinfo{volume}{67}
  (\bibinfo{year}{2018}), \bibinfo{pages}{94--105}.
\newblock
\urldef\tempurl%
\url{https://doi.org/10.1016/j.asoc.2018.02.051}
\showURL{%
\tempurl}


\bibitem[Mendon{\c{c}}a et~al\mbox{.}(2020)]%
        {DBLP:conf/icsa/MendoncaACG20}
\bibfield{author}{\bibinfo{person}{Nabor~C. Mendon{\c{c}}a},
  \bibinfo{person}{Carlos~Mendes Aderaldo}, \bibinfo{person}{Javier
  C{\'{a}}mara}, {and} \bibinfo{person}{David Garlan}.}
  \bibinfo{year}{2020}\natexlab{}.
\newblock \showarticletitle{Model-Based Analysis of Microservice Resiliency
  Patterns}. In \bibinfo{booktitle}{\emph{2020 {IEEE} International Conference
  on Software Architecture, {ICSA} 2020, Salvador, Brazil, March 16-20, 2020}}.
  \bibinfo{publisher}{{IEEE}}, \bibinfo{pages}{114--124}.
\newblock
\urldef\tempurl%
\url{https://doi.org/10.1109/ICSA47634.2020.00019}
\showDOI{\tempurl}


\bibitem[Methodology(2022)]%
        {ibm-resilience-test}
\bibfield{author}{\bibinfo{person}{IBM~Garage Methodology}.}
  \bibinfo{year}{2022}\natexlab{}.
\newblock \bibinfo{title}{Test software resiliency}.
\newblock
\newblock
\urldef\tempurl%
\url{https://www.ibm.com/garage/method/practices/manage/practice_resiliency/}
\showURL{%
\tempurl}


\bibitem[Morishita and Sese(2000)]%
        {morishita2000transversing}
\bibfield{author}{\bibinfo{person}{Shinichi Morishita} {and}
  \bibinfo{person}{Jun Sese}.} \bibinfo{year}{2000}\natexlab{}.
\newblock \showarticletitle{Transversing itemset lattices with statistical
  metric pruning}. In \bibinfo{booktitle}{\emph{Proceedings of the Nineteenth
  ACM SIGMOD-SIGACT-SIGART Symposium on Principles of Database Systems}}.
  \bibinfo{publisher}{Association for Computing Machinery},
  \bibinfo{pages}{226–236}.
\newblock
\showISBNx{158113214X}
\urldef\tempurl%
\url{https://doi.org/10.1145/335168.335226}
\showDOI{\tempurl}


\bibitem[Mostert et~al\mbox{.}(2018)]%
        {DBLP:conf/gecco/MostertME18}
\bibfield{author}{\bibinfo{person}{Werner Mostert},
  \bibinfo{person}{Katherine~M. Malan}, {and} \bibinfo{person}{Andries~P.
  Engelbrecht}.} \bibinfo{year}{2018}\natexlab{}.
\newblock \showarticletitle{Filter versus wrapper feature selection based on
  problem landscape features}. In \bibinfo{booktitle}{\emph{Proceedings of the
  Genetic and Evolutionary Computation Conference Companion, {GECCO} 2018,
  Kyoto, Japan, July 15-19, 2018}}. \bibinfo{publisher}{{ACM}},
  \bibinfo{pages}{1489--1496}.
\newblock
\urldef\tempurl%
\url{https://doi.org/10.1145/3205651.3208305}
\showDOI{\tempurl}


\bibitem[Neshatian and Zhang(2009)]%
        {DBLP:conf/gecco/NeshatianZ09}
\bibfield{author}{\bibinfo{person}{Kourosh Neshatian} {and}
  \bibinfo{person}{Mengjie Zhang}.} \bibinfo{year}{2009}\natexlab{}.
\newblock \showarticletitle{Pareto front feature selection: using genetic
  programming to explore feature space}. In \bibinfo{booktitle}{\emph{Genetic
  and Evolutionary Computation Conference, {GECCO} 2009, Proceedings, Montreal,
  Qu{\'{e}}bec, Canada, July 8-12, 2009}}. \bibinfo{publisher}{{ACM}},
  \bibinfo{pages}{1027--1034}.
\newblock
\urldef\tempurl%
\url{https://doi.org/10.1145/1569901.1570040}
\showDOI{\tempurl}


\bibitem[Newman(2015)]%
        {DBLP:books/lib/Newman15}
\bibfield{author}{\bibinfo{person}{Sam Newman}.}
  \bibinfo{year}{2015}\natexlab{}.
\newblock \bibinfo{booktitle}{\emph{Building microservices - designing
  fine-grained systems, 1st Edition}}.
\newblock \bibinfo{publisher}{O'Reilly}.
\newblock
\showISBNx{9781491950357}
\urldef\tempurl%
\url{https://www.worldcat.org/oclc/904463848}
\showURL{%
\tempurl}


\bibitem[Platt et~al\mbox{.}(1999)]%
        {platt1999probabilistic}
\bibfield{author}{\bibinfo{person}{John Platt} {et~al\mbox{.}}}
  \bibinfo{year}{1999}\natexlab{}.
\newblock \showarticletitle{Probabilistic outputs for support vector machines
  and comparisons to regularized likelihood methods}.
\newblock \bibinfo{journal}{\emph{Advances in large margin classifiers}}
  \bibinfo{volume}{10}, \bibinfo{number}{3} (\bibinfo{year}{1999}),
  \bibinfo{pages}{61--74}.
\newblock


\bibitem[Raza and Qamar(2016)]%
        {DBLP:conf/iotdcc/RazaQ16}
\bibfield{author}{\bibinfo{person}{Muhammad~Summair Raza} {and}
  \bibinfo{person}{Usman Qamar}.} \bibinfo{year}{2016}\natexlab{}.
\newblock \showarticletitle{A hybrid feature selection approach based on
  heuristic and exhaustive algorithms using Rough set theory}. In
  \bibinfo{booktitle}{\emph{Proceedings of the International Conference on
  Internet of Things and Cloud Computing, Cambridge, UK, March 22-23, 2016}}.
  \bibinfo{publisher}{{ACM}}, \bibinfo{pages}{47:1--47:7}.
\newblock
\urldef\tempurl%
\url{https://doi.org/10.1145/2896387.2896432}
\showDOI{\tempurl}


\bibitem[Samir and Kyle(2020)]%
        {resiliency-definition}
\bibfield{author}{\bibinfo{person}{Nasser Samir} {and} \bibinfo{person}{Brown
  Kyle}.} \bibinfo{year}{2020}\natexlab{}.
\newblock \bibinfo{title}{PRODUCTION SOFTWARE APPLICATION PERFORMANCE AND
  RESILIENCY TESTING}.
\newblock
\newblock
\urldef\tempurl%
\url{https://lens.org/094-261-625-935-507}
\showURL{%
\tempurl}


\bibitem[Services(2022)]%
        {aws-past-event}
\bibfield{author}{\bibinfo{person}{Amazon~Web Services}.}
  \bibinfo{year}{2022}\natexlab{}.
\newblock \bibinfo{title}{AWS Post-Event Summaries}.
\newblock
\newblock
\urldef\tempurl%
\url{https://aws.amazon.com/premiumsupport/technology/pes/}
\showURL{%
\tempurl}


\bibitem[Tipping and Bishop(1999)]%
        {pca}
\bibfield{author}{\bibinfo{person}{Michael~E. Tipping} {and}
  \bibinfo{person}{Christopher~M. Bishop}.} \bibinfo{year}{1999}\natexlab{}.
\newblock \showarticletitle{Mixtures of Probabilistic Principal Component
  Analysers}.
\newblock \bibinfo{journal}{\emph{Neural Comput.}} \bibinfo{volume}{11},
  \bibinfo{number}{2} (\bibinfo{year}{1999}), \bibinfo{pages}{443--482}.
\newblock
\urldef\tempurl%
\url{https://doi.org/10.1162/089976699300016728}
\showDOI{\tempurl}


\bibitem[Yang et~al\mbox{.}(2021)]%
        {AID}
\bibfield{author}{\bibinfo{person}{Tianyi Yang}, \bibinfo{person}{Jiacheng
  Shen}, \bibinfo{person}{Yuxin Su}, \bibinfo{person}{Xiao Ling},
  \bibinfo{person}{Yongqiang Yang}, {and} \bibinfo{person}{Michael~R. Lyu}.}
  \bibinfo{year}{2021}\natexlab{}.
\newblock \showarticletitle{{AID:} Efficient Prediction of Aggregated Intensity
  of Dependency in Large-scale Cloud Systems}. In
  \bibinfo{booktitle}{\emph{36th {IEEE/ACM} International Conference on
  Automated Software Engineering, {ASE} 2021, Melbourne, Australia, November
  15-19, 2021}}. \bibinfo{publisher}{{IEEE}}, \bibinfo{pages}{653--665}.
\newblock
\urldef\tempurl%
\url{https://doi.org/10.1109/ASE51524.2021.9678534}
\showDOI{\tempurl}


\bibitem[Yin and Du(2021)]%
        {DBLP:journals/ijseke/YinD21}
\bibfield{author}{\bibinfo{person}{Kanglin Yin} {and} \bibinfo{person}{Qingfeng
  Du}.} \bibinfo{year}{2021}\natexlab{}.
\newblock \showarticletitle{On Representing Resilience Requirements of
  Microservice Architecture Systems}.
\newblock \bibinfo{journal}{\emph{Int. J. Softw. Eng. Knowl. Eng.}}
  \bibinfo{volume}{31}, \bibinfo{number}{6} (\bibinfo{year}{2021}),
  \bibinfo{pages}{863--888}.
\newblock
\urldef\tempurl%
\url{https://doi.org/10.1142/S0218194021500261}
\showDOI{\tempurl}


\bibitem[Zhai et~al\mbox{.}(2020)]%
        {DBLP:conf/nsdi/ZhaiCPBTSZ20}
\bibfield{author}{\bibinfo{person}{Ennan Zhai}, \bibinfo{person}{Ang Chen},
  \bibinfo{person}{Ruzica Piskac}, \bibinfo{person}{Mahesh Balakrishnan},
  \bibinfo{person}{Bingchuan Tian}, \bibinfo{person}{Bo Song}, {and}
  \bibinfo{person}{Haoliang Zhang}.} \bibinfo{year}{2020}\natexlab{}.
\newblock \showarticletitle{Check before You Change: Preventing Correlated
  Failures in Service Updates}. In \bibinfo{booktitle}{\emph{17th {USENIX}
  Symposium on Networked Systems Design and Implementation, {NSDI} 2020, Santa
  Clara, CA, USA, February 25-27, 2020}},
  \bibfield{editor}{\bibinfo{person}{Ranjita Bhagwan} {and}
  \bibinfo{person}{George Porter}} (Eds.). \bibinfo{publisher}{{USENIX}
  Association}, \bibinfo{pages}{575--589}.
\newblock
\urldef\tempurl%
\url{https://www.usenix.org/conference/nsdi20/presentation/zhai}
\showURL{%
\tempurl}


\bibitem[Zhou et~al\mbox{.}(2021)]%
        {DBLP:journals/tse/ZhouPXSJLD21}
\bibfield{author}{\bibinfo{person}{Xiang Zhou}, \bibinfo{person}{Xin Peng},
  \bibinfo{person}{Tao Xie}, \bibinfo{person}{Jun Sun}, \bibinfo{person}{Chao
  Ji}, \bibinfo{person}{Wenhai Li}, {and} \bibinfo{person}{Dan Ding}.}
  \bibinfo{year}{2021}\natexlab{}.
\newblock \showarticletitle{Fault Analysis and Debugging of Microservice
  Systems: Industrial Survey, Benchmark System, and Empirical Study}.
\newblock \bibinfo{journal}{\emph{{IEEE} Trans. Software Eng.}}
  \bibinfo{volume}{47}, \bibinfo{number}{2} (\bibinfo{year}{2021}),
  \bibinfo{pages}{243--260}.
\newblock
\urldef\tempurl%
\url{https://doi.org/10.1109/TSE.2018.2887384}
\showDOI{\tempurl}


\bibitem[Zhou et~al\mbox{.}(2019)]%
        {DBLP:conf/sigsoft/Zhou0X0JLXH19}
\bibfield{author}{\bibinfo{person}{Xiang Zhou}, \bibinfo{person}{Xin Peng},
  \bibinfo{person}{Tao Xie}, \bibinfo{person}{Jun Sun}, \bibinfo{person}{Chao
  Ji}, \bibinfo{person}{Dewei Liu}, \bibinfo{person}{Qilin Xiang}, {and}
  \bibinfo{person}{Chuan He}.} \bibinfo{year}{2019}\natexlab{}.
\newblock \showarticletitle{Latent error prediction and fault localization for
  microservice applications by learning from system trace logs}. In
  \bibinfo{booktitle}{\emph{Proceedings of the {ACM} Joint Meeting on European
  Software Engineering Conference and Symposium on the Foundations of Software
  Engineering, {ESEC/SIGSOFT} {FSE} 2019, Tallinn, Estonia, August 26-30,
  2019}}. \bibinfo{publisher}{{ACM}}, \bibinfo{pages}{683--694}.
\newblock
\urldef\tempurl%
\url{https://doi.org/10.1145/3338906.3338961}
\showDOI{\tempurl}


\end{thebibliography}

\end{document}